\documentclass[fleqn,usenatbib]{mnras}

\usepackage{newtxtext,newtxmath}

\usepackage[T1]{fontenc}

\DeclareRobustCommand{\VAN}[3]{#2}
\let\VANthebibliography\thebibliography
\def\thebibliography{\DeclareRobustCommand{\VAN}[3]{##3}\VANthebibliography}

\usepackage[inline]{enumitem}

\setlist[enumerate]{align=left,leftmargin=8mm,labelsep*=0pt}
\setlist[enumerate,1]{label=(\alph*), ref=(\alph*)}
\setlist[enumerate,2]{label=(\roman*), ref=(\roman*),leftmargin=9mm}

\usepackage{graphicx}	
\usepackage{amsmath}	

\title[Precise radial velocities of double-lined binaries]{New methods for radial-velocity measurements of double-lined binaries, and detection of a circumbinary planet orbiting TIC 172900988}

\author[S. Lalitha  et al.]{Lalitha Sairam,$^{1,2}$\thanks{E-mail: lalitha.sairam@ast.cam.ac.uk}
Amaury H.M.J. Triaud,$^{1}$
Thomas A. Baycroft,$^{1}$
Jerome Orosz,$^{3}$
Isabelle Boisse,$^{4}$
\newauthor
Neda Heidari,$^{5}$
Daniel Sebastian,$^{1}$
Georgina Dransfield,$^{1}$
David V. Martin,$^{6}$
Alexandre Santerne,$^{4}$
\newauthor
Matthew R. Standing$^{7}$
\\
$^{1}$School of Physics and Astronomy, University of Birmingham, Edgbaston, Birmingham B15 2TT, UK\\
$^{2}$Institute of Astronomy, University of Cambridge, Madingley road, Cambridge CB3 0HA, UK\\
$^{3}$Department of Astronomy, San Diego State University, 5500 Campanile Drive, San Diego, CA 92182, USA \\
$^{4}$Aix Marseille Univ, CNRS, CNES, LAM, Marseille, France\\
$^{5}$Institut d'astrophysique de Paris, UMR 7095 CNRS université pierre et
marie curie, 98 bis, boulevard Arago,  75014, Paris\\
$^{6}$Department of Physics and Astronomy, Tufts University, 574 Boston Avenue, Medford, MA 02155\\
$^{7}$ School of Physical Sciences, The Open University, Milton Keynes, MK7 6AA, UK\\
}

\date{Accepted for publication in MNRAS}

\pubyear{2023}

\begin{document}
\label{firstpage}
\pagerange{\pageref{firstpage}--\pageref{lastpage}}
\maketitle

\begin{abstract}
Ongoing ground-based radial-velocity observations seeking to detect circumbinary planets focus on single-lined binaries even though over nine in every ten binary systems in the solar-neighbourhood are double-lined. Double-lined binaries are on average brighter, and should in principle yield more precise radial-velocities. However, as the two stars orbit one another, they produce a time-varying blending of their weak spectral lines. This makes an accurate measure of radial velocities difficult, producing a typical scatter of $10-15~\rm m\,s^{-1}$. This extra noise prevents the detection of most orbiting circumbinary planets. We develop two new data-driven approaches to disentangle the two stellar components of a double-lined binary, and extract accurate and precise radial-velocities. Both approaches use a Gaussian Process regression, with the first one working in the spectral domain, whereas the second works on cross-correlated spectra. We apply our new methods to TIC~172900988, a proposed circumbinary system with a double-lined binary, and detect a circumbinary planet with an orbital period of $150~\rm days$, different than previously proposed. We also measure a significant residual scatter, which we speculate is caused by stellar activity. We show that our two data-driven methods outperform the traditionally used TODCOR and TODMOR, for that particular binary system.

\end{abstract}




\section{Introduction}

Planets orbiting around both stars of a binary star system are called {\it p-type} or {\it circumbinary} planets. The discovery of Kepler-16 \citep{Doyle_2011} marked the first confirmation of the existence of these long-theorised planets  \citep{Borucki_1984, Schneider_1994}. Since then, the \emph{Kepler} and {\it TESS} missions have identified an additional 13 transiting circumbinary planets in 11 binary star systems \citep[e.g.][]{Martin2018, Kostov2020, Socia2020, Kostov2021}. While the radial velocity method is a highly effective way to detect exoplanets since \citet{Mayor1995}, only two circumbinary planets have been detected using this method: Kepler-16\,b \citep{triaud_2022} and TOI-1338/BEBOP-1\,c \citep{Standing2023}.

Ground-based radial-velocity surveys of binary stars are more efficient, and less biased than transit surveys \citep{martin_2019} and thus can construct a more insightful picture of the properties of the circumbinary planet population. The BEBOP survey \citep[Binaries Escorted By Orbiting Planets; ][]{martin_2019} began in 2017 as a dedicated, blind, radial-velocity survey of single-lined eclipsing binary stars. The survey reaches a precision of a few metres per second \citep{triaud_2022}. However, the BEBOP survey currently remains confined to highly unequal stellar-mass pairs where the secondary's mass is usually less than 30\% of the primary star \citep{Triaud_2017, martin_2019}. These single-lined binaries represents only around 8\% of all binaries, however, the remaining 92\% of binaries are double-line binaries \citep{Kovaleva_2016}. Prior to the BEBOP survey, an extensive survey of double-line binaries was carried out by the TATOOINE survey \citep{Konacki_2009, Konacki_2010}.
However, TATOOINE did not yield any circumbinary planets and reported a $10-15~\rm m\,s^{-1}$ scatter in the data, enough to hide most exoplanets. \citet{Konacki_2009} had to first deconvolve their spectra to remove an iodine-cell absorption spectrum, then they use TODCOR \citep{zucker_1994} to obtain a guess radial-velocity for each component of the binary. Then they perform their tomographic deconvolution method to accurately  measure the radial-velocity for each of the stellar components. 

Radial velocities are also often extracted by cross-correlating the observed spectrum with a template spectrum, or with a line-list mask \citep[the method we use; e.g.][]{Baranne_1996}. When the template matches the observed spectrum a strong signal is recorded.  The cross-correlation method is convenient to understand what might help address radial velocity scatter.

\begin{figure*}
\includegraphics[width=0.49\textwidth]{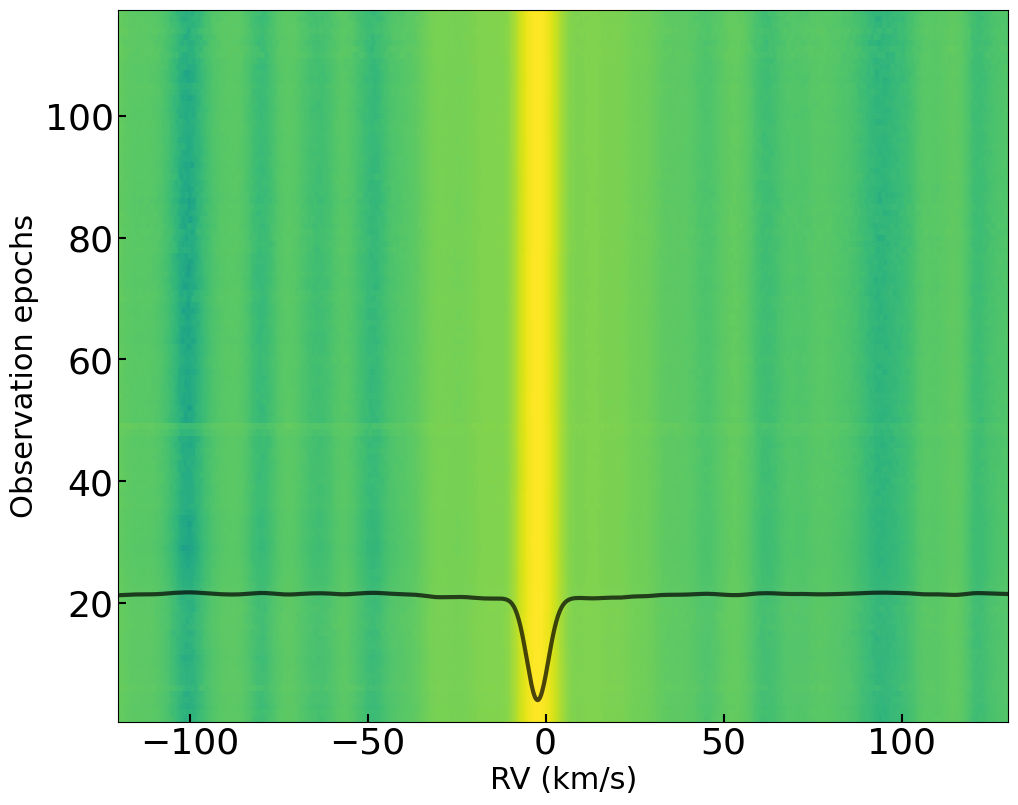}
\includegraphics[width=0.4375\textwidth]{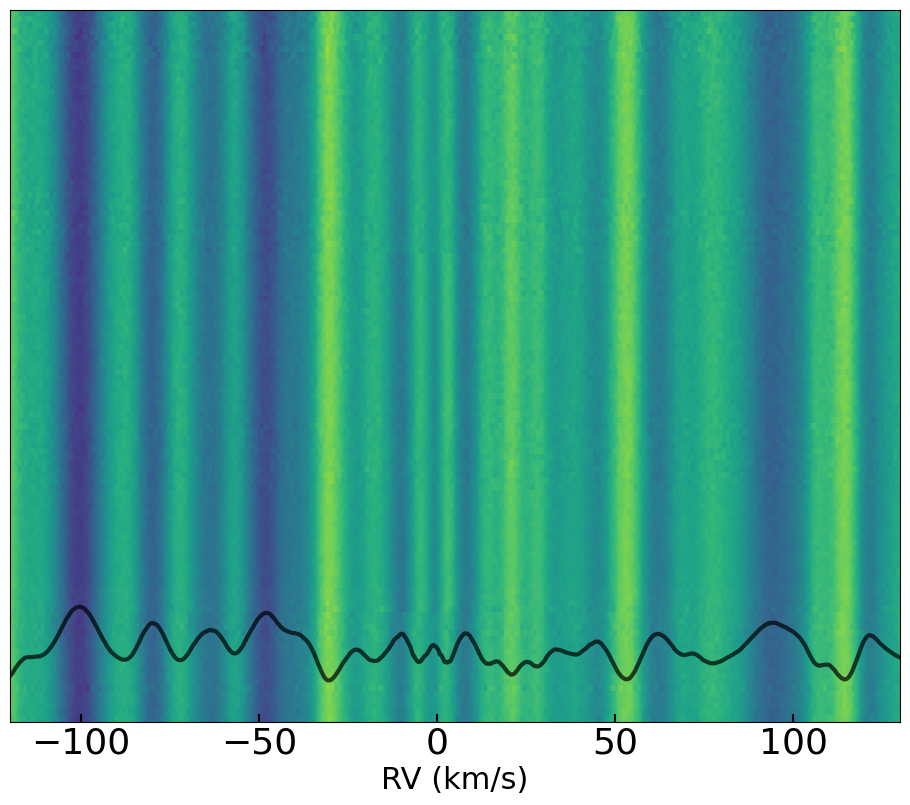}
\caption{Left panel: A time series of cross correlation function for HD 189733. The bright profile shows the stellar profile. The wiggles appear on either side of bright profile. The black profile is the median CCF of all epochs. Right panel: Residual map of time series on removal of stellar profile. Wiggles are seen as alternating static bands. Overlaid in black is the median wiggle of all epochs.}
\label{fig:singlewiggle}
\end{figure*}

In addition to the large cross-correlation function (CCF) signal, weaker signals on either side of the main signal are also recorded. They are caused by coincidental correspondence between the template and the spectrum, which we call {\it wiggles}. The wiggles exhibit pseudo-static behaviour over time in relation to the main CCF signal. We refer to these wiggles as pseudo-static because their characteristics may vary depending on the observation conditions. As an example, Figure \ref{fig:singlewiggle} depicts a CCF time-series showing the strong stellar signal for HD\,189733 but also its weaker, and stable wiggle signals.

\begin{figure}
    \centering
    \includegraphics[width=0.48\textwidth]{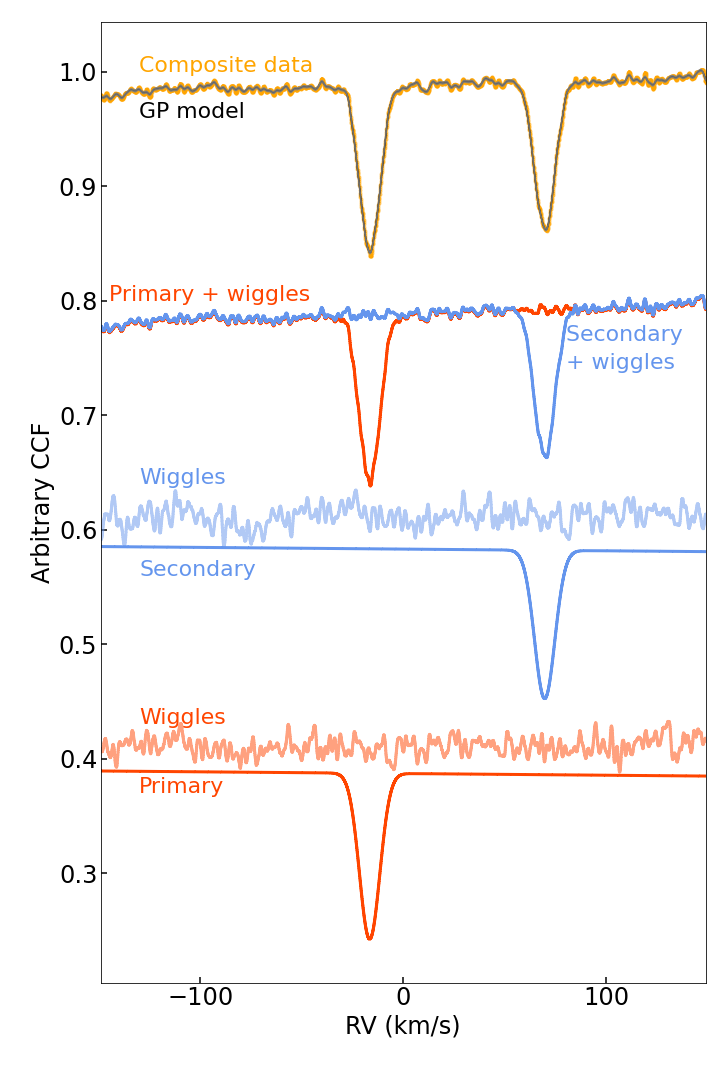}
    \caption{Schematic diagram showing the double-line binary problem and our solution to the problem using method 2 (CCF-GP)}
    \label{fig:ccf_scheme}
\end{figure}

 In the context of double-line binary stars, the presence of two bright stars leads to two distinct, and strong CCF signals that vary in time based on the respective masses and orbital parameters of the binary system. Figure \ref{fig:ccf_scheme} illustrates this time-varying signal, as well as the accompanying wiggles. Because there are two stars, there are two sets of wiggles. Each can interact with the strong CCF signal of the other star, and blending of a wiggle from star A with the CCF of star B, can lead to an error in estimating the radial-velocity of star B (and vice-versa). This effect also happens on the spectral side, where weak lines from star A can blend with strong lines of star B. If the template used for deconvolution is incomplete, this will cause a similar issue when estimating radial-velocities. This phenomenon, which we refer to as the {\it double-lined binary problem}, is likely the issue that prevents the discovery of circumbinary planets. 
 
 With this effect in mind, it is obvious that a data-driven approach needs to be taken since no template can properly reproduce every spectral feature. Any error on the spectrum/template spectrum of one star, can affect the radial-velocity measurement of the other star.

In this paper, we present two new methods to derive precise radial velocities of double-lined binaries. We treat the wiggles as a correlated signal. Such correlated signals are typically and accurately treated in a data-driven way using Gaussian processes \citep{aigrain_2016}. This method has had several successes \citep[e.g.][]{Czekala_2017, Rajpaul_2020} within the exoplanet field.

The structure of the paper is as follow: \S\ref{sec:gpintro} we introduce the Gaussian process (GP) framework. We then describe the two new techniques we developed for deriving radial velocities of double-lined binary stars. In \S\ref{sec:applygp}, we  choose to apply our two methods to the double-lined binary system TIC\,172900988  because of the presence of a planet in the system that allows to test for the recovery of a Keplerian signal. In \S\ref{sec:results}, we describe how we analyse the resulting radial velocities to model a binary's Keplerian motion, search for a circumbinary planet and infer its orbital parameters. Finally, we compare the radial velocities derived from our two new methods, and compare them to traditional and publicly available methods of measuring radial velocities \citep[TODCOR and  TODMOR;][]{zucker_1994,Mazeh_1994,zucker_2004}. 
We also discuss the implication of detecting  a circumbinary planet in TIC\,172900988 binary system. We conclude the paper with a summary of our key results in \S\ref{sec:conclusion}.

\section{Gaussian process-based radial velocity extraction}\label{sec:gpintro}

In this section, we provide a brief overview of the GP framework, which is a non-parametric Bayesian modelling technique that we use to infer the spectra of double-lined binary star systems. A GP is a type of stochastic process that describes the distribution of a group of random variables. It can be thought of as an extension of kernel regression to probabilistic models. Using non-parametric GPs to model the unknown wiggle function of double-lined binaries is a powerful and flexible approach \citep{rasmussen_2006}. 

A normal distribution is often represented as $\mathcal{N}(\mu,\sigma^2)$. If a random variable $x$ is normally distributed with mean and variance, it can be expressed as 
\begin{equation}\label{eqn:gp1}
    x \sim \mathcal{N}(\mu_x,\Sigma_{xx})
\end{equation}
with $\mu_x$ represents mean vector and $\Sigma_{xx}$ represents the covariance matrix. 
The covariance matrix describes the pairwise covariance between the different elements of the input data $x$. 

The likelihood that a set of observations $y$ is drawn from GP can be written as 
\begin{align}
    \ln \mathcal{L} &= \ln p(y|x, \phi,\theta)\\
    &= -\frac{1}{2}(y-\mu)^{\mathrm{T}}\, K^{-1} (y-\mu) - \frac{1}{2}~\ln |K| - \frac{N}{2}\ln(2\pi), \label{eq:logLike}
\end{align}
where $\phi$ and $\theta$ are hyperparameters of the mean and covariance functions. In Eq.~\ref{eq:logLike}, $K$ refers to the covariance matrix associated with the GP. The elements of the covariance matrix depend on the chosen covariance function and the values of the hyperparameters $\theta$. $N$ represents the number of elements in the vector $y$, which contains the observations. Evaluating this likelihood provides a posterior distribution of the hyperparameters.

\begin{figure}
\includegraphics[width=0.45\textwidth]{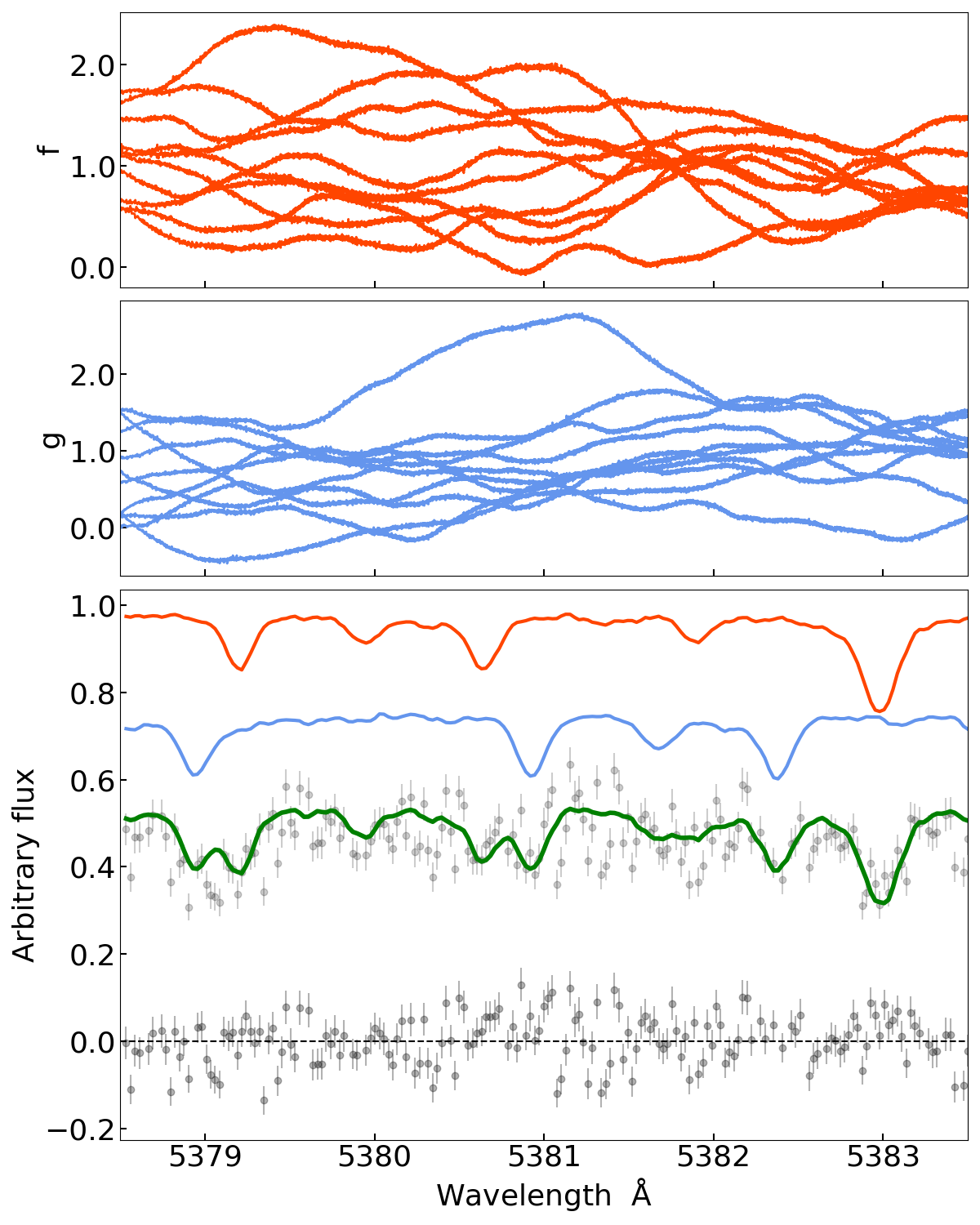}
\caption{Depicted in the top two panels are the multivariate realisation vector from prior distribution draw for GP hyperparameters of the primary and the secondary stars. The realisations are not constrained by the data. The bottom panel depicts the mean prediction (in green) from the posterior predictive distribution of GP drawn by conditioning on the observed data from a single epoch observation. The orange and blue in the bottom panel represents the primary and the secondary star with an arbitrary offset. The bottom panel also includes the residuals.}
\label{fig:sdillus}
\end{figure}

\begin{figure*}
\includegraphics[width=0.48\textwidth]{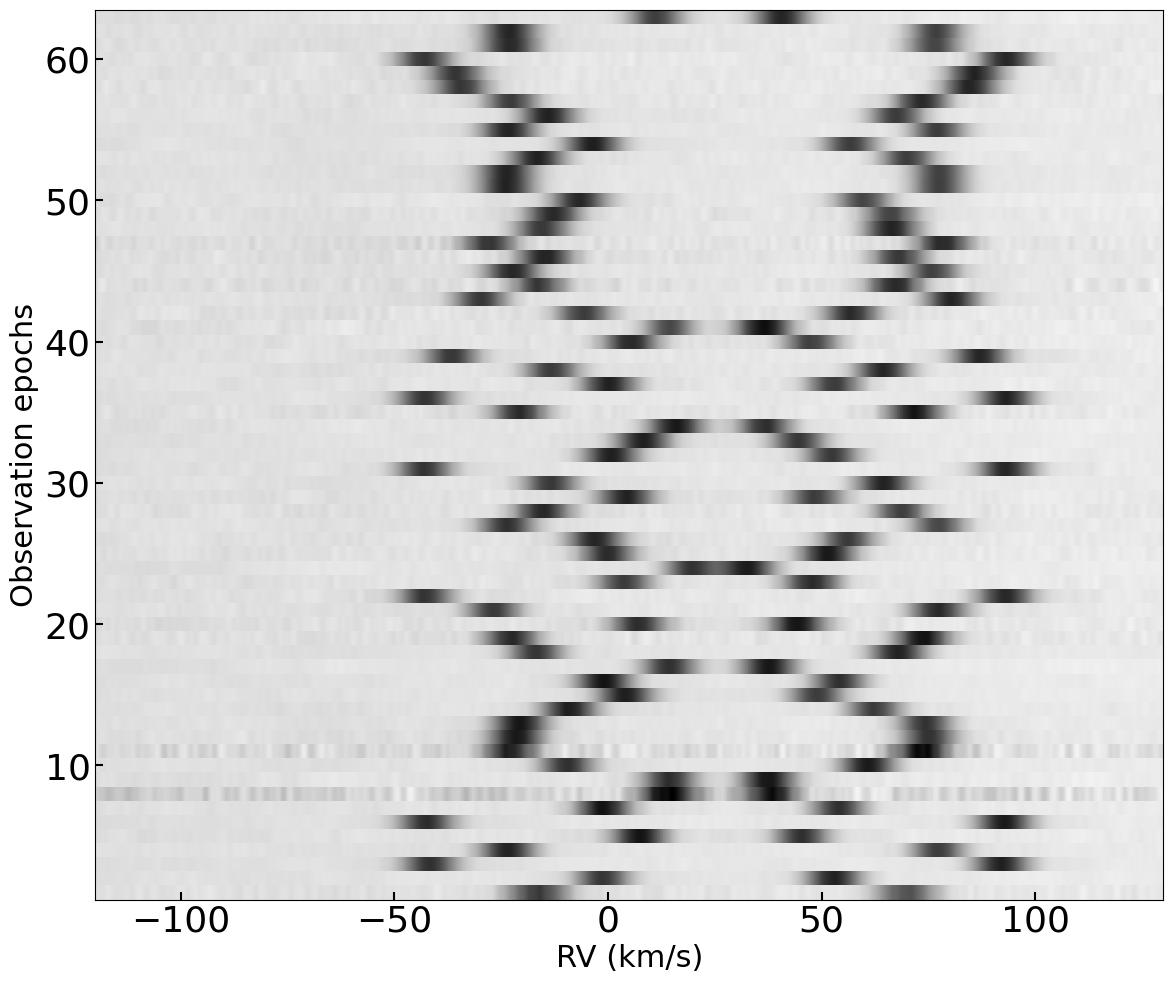}
\includegraphics[width=0.48\textwidth]{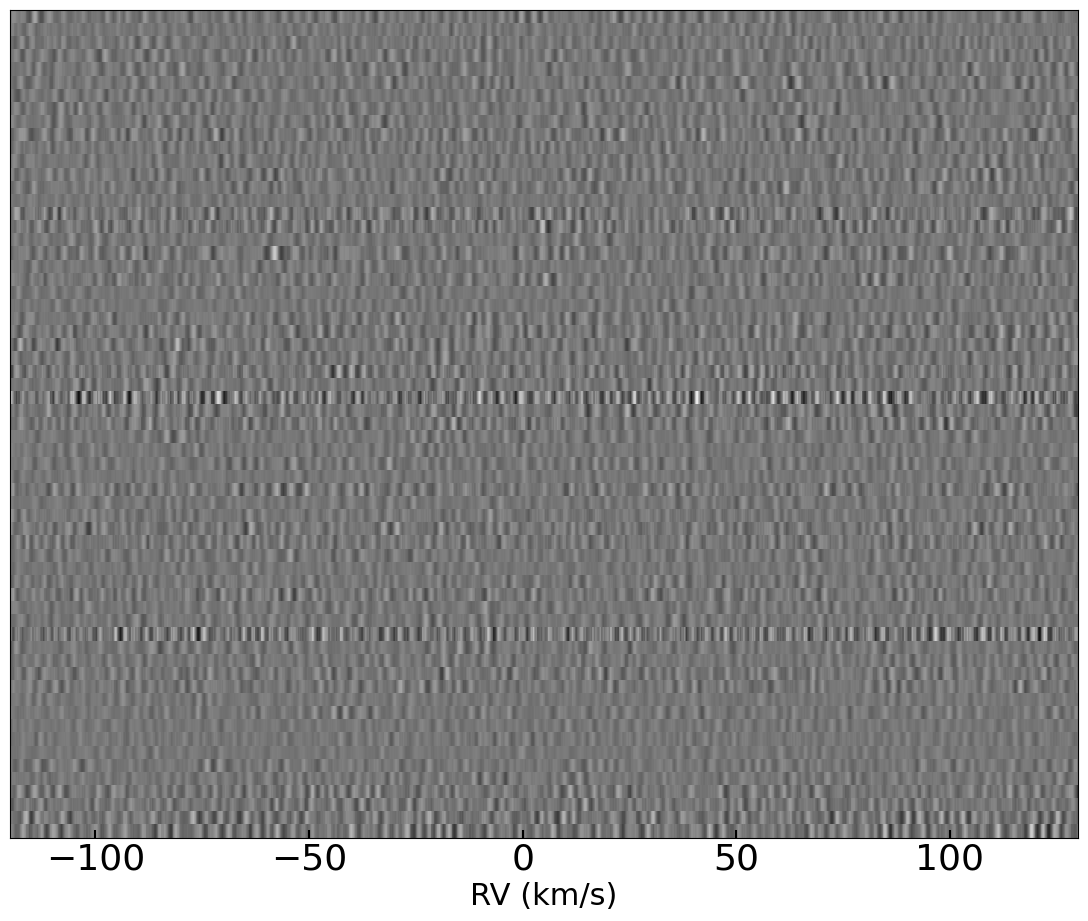}
\caption{ Left panel: A time-series of cross correlation for double-line binary star showing the strong signals for each components. The wiggles appear as dark bands on either sides of the strong signal. Right panel: The wiggles  as modelled by method 2 (CCF-GP).} 
\label{fig:ccf_binary_approach2}
\end{figure*}

\subsection{Method 1: Efficient Spectral Decomposition using Gaussian Processes (SD-GP)}\label{sec:sd-gp}

The observed spectra of a double-line binary star system can be modelled as a GP \citep[e.g.][]{Czekala_2017}. The spectrum of a single star can be represented as a function $f(\lambda$) where $\lambda$ is the wavelength. If $\lambda > 0$, the observed spectrum of a single star can be modelled as a function $f(\lambda)$ with a mean function $\mu(\lambda)$ and a covariance kernel $k(\lambda,\lambda')$. If the observed spectrum has finite inputs $0< \lambda_1 < \lambda_2 < ......< \lambda_w$, then the vector $\left[{f(\lambda_1), f(\lambda_2),.....f(\lambda_w)}\right]$ has a multivariate Gaussian distribution with a mean function $\left[{\mu(\lambda_1), \mu(\lambda_2), .....\mu(\lambda_w)}\right]$ and a covariance matrix with $k(\lambda_i,\lambda_j)$ as its elements, where $k$ is the kernel function and $i,j = 1,2, ..., w$. Therefore, the intrinsic continuous spectrum of a star can be assumed to be a function $f$ generated from a GP:
\begin{equation}
    f(\lambda)\sim {\rm GP} \left({\mu(\lambda), k(\lambda,\lambda')}\right).
\end{equation}

The observed spectra of a spectroscopic binary are composed of spectral lines from both stars as a composite. Due to the orbital motion of each stars in the binary, a Doppler shift is induced to the rest frame stellar spectra which are observed in the composite spectra simultaneously.
For a star moving with radial velocity \emph{v}, the rest frame wavelength of the observed spectra are shifted to 
\begin{equation}
    \lambda(v) = \left(\frac{c+v}{c-v}\right)^{\frac{1}{2}} \lambda_0,
\end{equation}
where $c$ is speed of light. 

We describe the radial velocities of binary stars as a function of time using seven parameters: the semi-amplitude of the primary star ($K_{\mathrm{A}}$), the binary stars mass ratio  $\left(\mathrm{q} = \frac{M_{\mathrm{B}}}{M_{\mathrm{A}}}=\frac{K_{\mathrm{A}}}{K_{\mathrm{B}}}\right)$,  binary orbital period ($P$), the eccentricity ($e$), the argument of periastron ($\omega$), the epoch of periastron ($T_0$) and systemic velocity ($\gamma$). The velocity of the primary and secondary stars as a function of time is 
\begin{equation}
    v_A = K_A [\cos(\omega+f(t)) + e \cos\omega] + \gamma
\end{equation}
and 
\begin{equation}
    v_B = -\frac{K_A}{q} [\cos(\omega+f(t)) + e \cos\omega] + \gamma,
\end{equation}
respectively. 

For a single epoch observation of a double-line binary star, we assume that the observed composite spectrum ($s$) is a sum of realisation $f$ for the primary star and $g$ for the secondary star 
along with $N$, the noise process realisation. 

\begin{align}
    s&= f + g + N\\
    &\sim \mathcal{N}(\mu_f,\Sigma_{f})  + \mathcal{N}(\mu_g,\Sigma_{g}) + \mathcal{N}(0,\Sigma_{N})\\
    &\sim \mathcal{N}(\mu_f+\mu_g,\Sigma_{f}+\Sigma_{g} + \Sigma_{N})
\end{align} 
where $\Sigma_{f}$ and $\Sigma_{g}$ are covariance matrices describing the primary and secondary star. We evaluate the covariance matrix $\Sigma_{f}$ and $\Sigma_{g}$ using the wavelengths corresponding to the primary and secondary components in their rest frame with the kernel function (Eq. \ref{eq:kernel}).   In Figure \ref{fig:sdillus} we show the realisation $f$ and $g$ for both components of the binary star.

Similar to \citet{Rajpaul_2020}, we choose the Mat\'ern kernel to model our spectra. The Mat\'ern kernel is often used when modeling spectra because it is a flexible and versatile covariance function that can capture a wide range of smoothness and correlation properties.
The Mat\'ern covariance kernel specifies the covariance between two pixels $\lambda_i$ and $\lambda_j$ as 
\begin{equation}\label{eq:kernel}
    k_{ij}(r_{ij}|\theta) = \sigma_f^2 \left(1+ \frac{\sqrt{3}r}{\sigma_l} \right) \exp\left(-\frac{\sqrt3 r}{\sigma_l}\right)
\end{equation}
where r$_{ij}$ has units of km s$^{-1}$

\begin{equation}
    r_{ij} = r(\lambda_i,\lambda_j) = \frac{c}{2} \left|\frac{\lambda_i-\lambda_j}{\lambda_i+\lambda_j}\right|,
\end{equation}

with $c$ as the speed of light, $\sigma_l$ the characteristic length scale and $\sigma_f$ the signal standard deviation.

In practice, we analyse our spectra with the following steps and assumptions:
\begin{enumerate}[labelwidth=0.4cm]
    \item The observed composite spectra are divided into smaller wavelength subsets, which we call {\it chunks}. The radial velocity shift is computed separately for each chunk. In principle this step is not necessary, but GPs are computationally intensive and without breaking each spectrum into smaller components, the calculation become intractable. 
    \item We allow separate values for GP hyperparameters $\sigma_f$ and $\sigma_l$ (i.e., $\{\sigma_f,\sigma_l\}_f$,$\{\sigma_f,\sigma_l\}_g$) for each star of the binary system. This allows an optimal reconstruction of the spectrum taking into account the different spectral types of the stars. The mean function that is obtained by drawing samples from the GP distribution is shown in Figure \ref{fig:sdillus} (bottom panel).
    \item Rather than forcing a single set of hyperparameters to model a spectrum, we allow different sets of hyperparameters for different regions of a spectrum.
    \item To set reasonable initial values for the radial velocities of the primary and secondary stars, we first fit simple Gaussian functions to the cross-correlation functions of each component. We then use that result to define a flat/uniform prior distribution on the radial velocities of each component.
    A flat prior distribution assigns equal probability density to all values within a specified range. The bounds of the flat prior distribution can be defined as the minimum and maximum values of the range, which would depend on the expected range of radial velocities for each component. We chose the bounds carefully to avoid assigning unrealistic probabilities to certain values.
    We then use a $\chi^2$ likelihood function, to refine these estimates.
    \item We simultaneously explore the posterior distribution of the radial velocities and the GP hyperparameters using an MCMC (Markov Chain Monte Carlo) sampler \citep[the {\sc emcee} python package;][]{goodman2010ensemble, Foreman-Mackey_2013}. 
    \item We reiterate these steps for each chunk, and then filter bad radial velocity estimates caused by telluric absorption, stellar activity contamination or instrumental systematics. 
    \item The radial velocities from individual chunks are combined by computing a weighted average. The weights used for computing the weighted average of the radial velocities from individual chunks are determined by the uncertainties associated with the radial velocity estimates obtained from each chunk. 
     \item We normalise the observed spectra, and for the GP, we set the mean function to a constant value of 1.0 ($\mu$=1).   
\end{enumerate}

\subsection{Method 2: Cross-Correlation Functions modelled using Gaussian process (CCF-GP)}\label{sec:ccf-gp}

Method 1 involves dividing the observed spectrum covering a large wavelength range into smaller chunks and applying a GP regression on each chunk. This process involves computing the kernel matrix, inverting it, and multiplying it by the training set data. The computational complexity of GP regression is $\mathcal{O}(N^3)$, where $N$ is the number of data points in each chunk. Therefore, the overall complexity of method 1 would be $\mathcal{O}(MN^3)$, where $M$ is the number of chunks in the spectrum. This can be computationally very expensive. Hence we also develop an alternative method.

This alternative approach is similar to the previous one but instead of modelling the entire spectrum chunk by chunk, we instead model the cross-correlated spectra, which are a typical output of instruments such as HARPS, SOPHIE and ESPRESSO \citep{Baranne_1996,Pepe_2002,Perruchot_2008}. 
We assume that the cross-correlation functions (CCFs) are 
samples of GP. The mean function ($\mu(x)$) is constructed as the sum of two Gaussian functions. The covariance kernel function is used to model the correlated wiggle signal found within the cross-correlation function.

We employ a Gaussian fit jointly with a GP model for each of the components of binary. 
The baseline mean-function is a Gaussian function for each of the component of the binary 

\begin{equation}
    \mu(x) = 1 - \mathrm{A}_1 \exp\left( -\frac{(x - \mathrm{B}_1)^2}{2\mathrm{C}_1^2}\right) - ~ \mathrm{A}_2  \exp \left( -\frac{(x - \mathrm{B}_2)^2}{2\mathrm{C}_2^2}\right)
\end{equation}
where A$_1$, B$_1$, C$_1$, A$_2$, B$_2$, C$_2$ are free hyper-parameters. A$_1$ and A$_2$ correspond to the amplitude of the Gaussian, which represents the contrast of primary and secondary components. B$_1$ and B$_2$ represent the radial velocities of primary and secondary stars, while C$_1$ and C$_2$ correspond to the standard deviation of the Gaussian, which represents the full width half maximum (FWHM).

We create a custom model class that inherits from {\it celerite.modeling.Model}, which is used to define the mean function of the GP.
We use a Mat\'ern covariance kernel (Equation \ref{eq:kernel}), implemented in \texttt{celerite} \citep{Foreman-Mackey_2017} to model the correlated wiggle signal. We set bounds on the input parameters for the model and the hyperparameters of the Mat\'ern kernel, and then creates two Mat\'ern kernels, one for each Gaussian component, which are combined into a single kernel.

We optimise the GP model using the L-BFGS-B method \citep{Byrd_1995}, which allows us to impose bounds on the parameters while minimising the negative log-likelihood of the model. We use MCMC sampling to explore the posterior distribution of the hyperparameters. For this we also use the {\sc emcee} python package \citep{goodman2010ensemble, Foreman-Mackey_2013} with 50 walkers and a burn-in of 100 iterations. We set broad uniform priors for each hyperparameter, and run the final MCMC with 5000 iterations to converge on a solution. We take the median of the posterior distribution as the optimum solution for each hyperparameter.  We then  compute the 16th, 50th (median), and 84th percentiles of the posterior distribution. The uncertainties of the hyperparameters are taken as the difference between the 84$^{\mathrm{th}}$ and 50$^{\mathrm{th}}$ percentile (upper bound) and 50$^{\mathrm{th}}$ and 16$^{\mathrm{th}}$ percentile (lower bound).

The most intensive part of this method is the optimisation of GP hyperparameters which is performed by L-BFGS-B method. However, for this method the main computational bottleneck could  likely be the MCMC sampling step, which has a complexity that scales with the size of the CCF.  

\section{Application  to TIC~172900988}\label{sec:applygp}

TIC~172900988 is an eclipsing double-lined binary system consisting of two stars with spectral types F9 and G0. The orbital period of the system is approximately $\sim$19.7 days. 
\cite{Kostov2021} reported the first discovery of a circumbinary planet via what is sometimes called the "1-2 punch technique", where multiple transits occur during one conjunction event, the planet transits once over the primary and once over the secondary. \cite{Kostov2021} used a photo-dynamical analysis but did not find a single solution. The planetary radius is constrained at ${\rm R}_{\rm p} = 11.25\pm0.44~\rm R_\oplus$. The planetary masses are proposed within a range of $823 < m_{\rm p} < 981~\rm m_\oplus$, and the orbital period within $188< \rm P_{\rm p} < 204~\rm d$. 

We have collected radial-velocities with SOPHIE on 10 double-line binaries \citep[including six from ][]{Konacki_2009,Konacki_2010} in order to test our methods to different spectral types, orbital solutions, relative velocities, etc, and observe their limitations. Given the presence of a planet within the TIC~172900988 system, it serves as a good first testbed to demonstrate our ability to extract radial-velocities without removing a Keplerian signal. By focusing first on a known circumbinary planet host, we can evaluate the effectiveness and accuracy of our approaches. We plan to follow this paper with another paper analysing the rest of the sample that will show how precise our new methods are.

\subsection{Observation}\label{section:obs}

We collected 62 epochs of high-resolution spectra between 2020 October 16 and 2023 May 05 using the SOPHIE spectrograph mounted on the 1.93m telescope at Observatoire de Haute Provence (OHP) in France \citep{Perruchot_2008}. The spectra cover a wavelength range of 3872-6943 $\Angstrom$ in 39 spectral orders, with a resolving power of $\delta\lambda/ \lambda \approx75,000$. The exposure times ranged from 600 to 1800~s depending on the seeing conditions at OHP. They have an median signal to noise ratio of $32$ at 5500 $\Angstrom$. 
These are signal to noise ratio for the composite spectra of the TIC~172900988 with an average flux fraction of 0.86. This corresponds to an signal to noise ratio of $\sim$17 and $\sim14$ at 5500 $\Angstrom$ for the primary and the secondary, respectively. SOPHIE was designed to detect exoplanets with a long-term stability of $2~\rm  m\,s^{-1}$. The observations were taken in objAB mode, where one fibre is used to observe the starlight and another fibre is used to observe the sky brightness to estimate the background contamination such as that produced by moonlight. The wavelength calibration was performed before the night starts using a Thorium-Argon lamp and a Fabry-Pérot, fed into both fibres. Additional Fabry-Pérot calibrations are obtained roughly every two hours within the night. The spectra are extracted using the SOPHIE automatic pipeline \citep{Bouchy_2009} and the resulting wavelength-calibrated spectra are correlated with a numerical binary mask to obtain the cross-correlation functions \citep{Baranne_1996, Pepe_2002}. We used a G2 mask for the correlation.

\subsection{Method 1 -  SD-GP}\label{sec:sd-gp2}

We first obtain the spectra and and cross-correlate them using the SOPHIE Data Reduction Software. 
To effectively apply Method 1, we work with two dimensional spectra (e2ds) at the instrument resolution instead of using 1D spectra (s1d), which operate at the pixel sampling level. We measure the radial velocities of both stellar components at the time of observation using a Gaussian fit to their cross-correlated spectra. Each SOPHIE spectrum covers from 3872 $\Angstrom$ to 6943$\Angstrom$. We divide each observed spectrum into chunks of $5\Angstrom$ each, totalling to 615 chunks. 
For each chunk, we apply the SD-GP method to measure the radial velocities of both stars at each epoch. 
Using the calculated velocities and the parameters of the GP, we deconvolve the composite spectra into the individual spectrum of both individual stars for each epoch, by optimising the parameters of the model to fit the observed spectra. In Figure~\ref{fig:sdphase} (left panels), we give examples of this step of our analysis, where we show median of posterior predictive distribution of the predicted composite spectrum and the reconstructed spectra of each component of the binary. For better visualisation, we have arbitrarily included an offset to each spectrum. Note that the reconstructed spectrum matches the shape of the input composite spectrum. It is important to note that there may be chunks where the spectral lines are not present, possibly due to the continuum dominating the spectrum (Figure~\ref{fig:sdphase}, right panels), resulting in large uncertainties in the RV measurements. After repeating this process for each chunk of spectra, we obtain the radial velocities for each star in the binary system at each epoch. We then apply outlier removal using a Student's-t distribution to remove any radial velocities that lie outside of the 95\% confidence interval. The remaining radial velocities are assigned weights considering the associated uncertainties to estimate the weighted average radial velocities for the binary system (see \S\ref{sec:sd-gp}). We then estimate the uncertainty of the combined radial velocity by propagating the uncertainties of each individual chunk through the weighted average.

\subsection{Method 2 - CCF-GP}\label{sec:ccf-gp2}

All 62 epochs of spectroscopic data from SOPHIE were cross-correlated with a  G2 mask. To determine the radial velocities of the primary and secondary star in the binary system TIC~172900988, we apply the CCF-GP to the resulting CCFs. To measure the radial velocities of the primary and secondary stars, we fit the cross-correlation function with a double Gaussian function, with the two Gaussians representing the primary and secondary stars, and two GPs to model each of {\it wiggles} caused by coincidental correspondence with the mask. In Figure \ref{fig:ccf_binary_approach2} (left panel), we show the cross-correlation function time series along with the wiggles. We explore the posterior distribution of hyperparameters using MCMC sampling. We then calculate the 16$^{\mathrm {th}}$, 50$^{\mathrm {th}}$, and 84$^{\mathrm {th}}$ percentiles of the samples of the radial velocity, FWHM and contrast  obtained from the MCMC simulation. These percentiles correspond to the lower uncertainty limit, the median value, and the upper uncertainty limit of the hyperparameters, respectively. The residuals following the subtraction of each component of the binary and the wiggles are shown in Figure \ref{fig:ccf_binary_approach2} (right panel).

\begin{figure}
\includegraphics[width=0.49\textwidth]{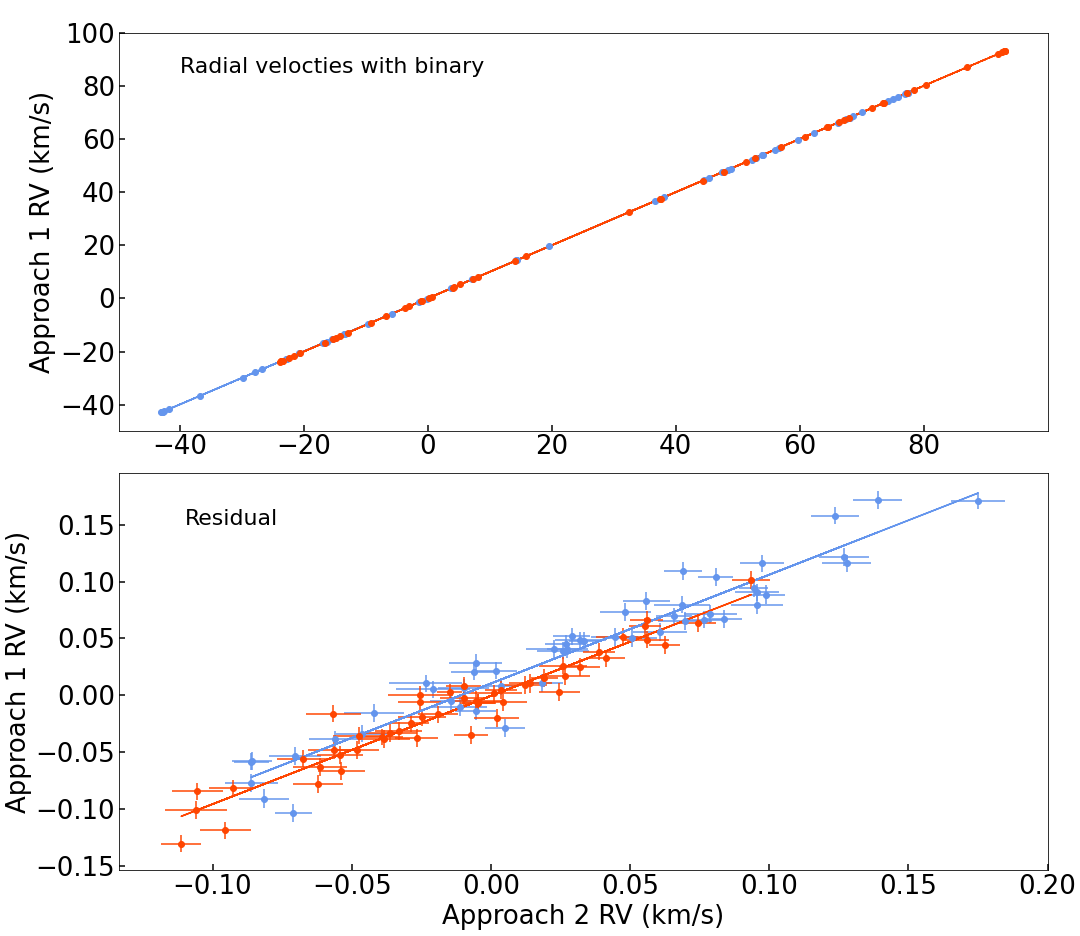}
\caption{Top panel: Radial velocities extracted using method 1 versus the radial velocities from method 2, along with 1$\sigma$ error bars. The red and blue colours represent the primary and the secondary stars along with their respective 1:1 identity lines. Bottom panel: The residual radial velocities after removing the binary signal for method 1 versus the residual radial velocities for method 2.}
\label{fig:comp2}
\end{figure}

\subsection{TODCOR and TODMOR}\label{sec:todcor}
We also compute radial-velocities using traditional methods such as Two-Dimensional Correlation (TODCOR) and TODMOR (Two-Dimensional Modelling and Reconstruction). TODCOR  is a method that uses a two-dimensional cross-correlation (2D-CCF) to measure the radial velocities of the primary and secondary stars in a binary system \citep{Mazeh_1994, zucker_1994}. 
A modern implementation of TODCOR for multi-order spectra is TODMOR \citep{zucker_2004}. TODMOR also uses a two-dimensional model of the observed spectra. To measure the radial velocities of the primary and secondary stars TODMOR compares each stellar component with a template spectrum matching their spectral type.

In the next section, we compare our two new methods to results produced by TODCOR and TODMOR. As such we apply TODCOR and TODMOR to the same observed spectra that we used for our own approaches. To apply TODCOR and TODMOR, first we need to correct the SOPHIE spectra for the instrumental blaze function, and detrend the pseudo-continuum. We then use the PHOENIX stellar model \citep{husser_2013} to determine the best-matching theoretical spectra for the primary and secondary stars, and use these to optimise the 2D-CCF for each order of the spectra. We apply TODMOR to each SOPHIE order and determine the radial velocities of both components, discarding orders strongly influenced by telluric lines.

TODCOR and TODMOR use template spectra for primary and secondary stars and cross-correlate them to the observed composite spectrum to determine the radial velocities. Neither TODCOR nor TODMOR treat the wiggles.  

\section{Results}\label{sec:results}

\subsection{Binary model}

\begin{figure*}
\includegraphics[width=\textwidth]{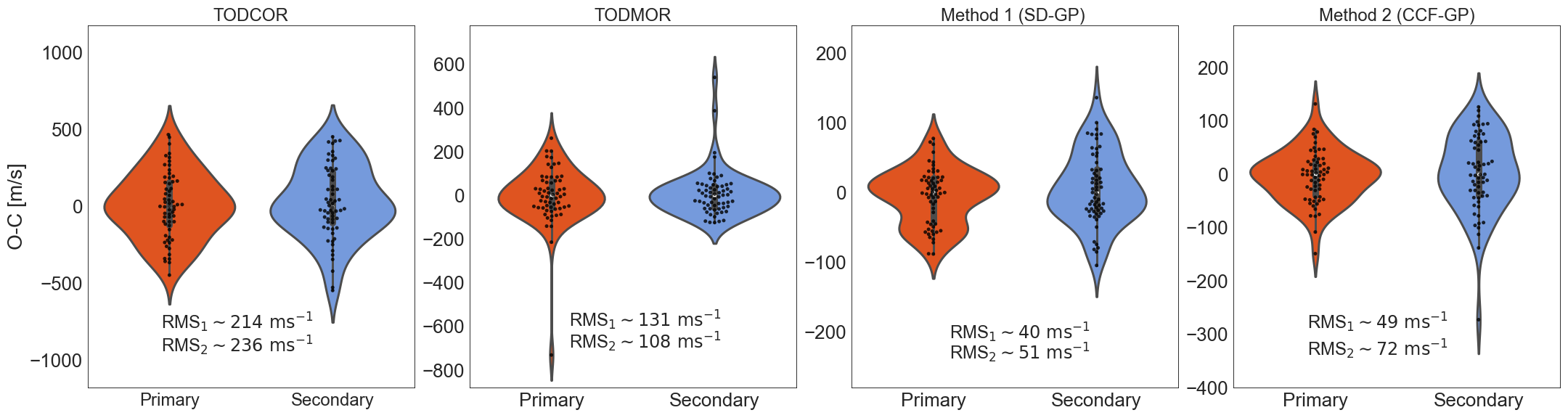}
\caption{The distribution of O-C values for the primary (red) and secondary stars (blue) in TIC172900988, as determined by TODCOR, TODMOR, method 1 (SD-GP), and method 2 (GP-CCF), are displayed using violin plots. Note that the vertical scale is different on every panel.}
\label{fig:compallrms}
\end{figure*}

Each of the measured radial velocities obtained from method 1 (Sec.~\ref{sec:sd-gp} and \ref{sec:sd-gp2}), method 2 (Sec.~\ref{sec:ccf-gp} and \ref{sec:ccf-gp2}), TODCOR, and TODMOR (Sec.~\ref{sec:todcor}) are independently modelled. We utilise \texttt{kima}, an open source software package for fitting radial velocities, to determine the physical parameters of the binary \citep{faria_2018}. Specifically, an updated \texttt{kima} package is used, which now includes simultaneous fitting of both components of double-lined binaries, correction for General Relativity effects, and that can fit for apsidal precession of the binary \citep{Baycroft_2023}. For sampling, \texttt{kima} employs a diffusive nested sampling algorithm \citep[DNest4,][]{brewer_dnest4_2018}. To account for stellar variability effects, a radial velocity jitter term is incorporated. Outliers are included in the procedure, and are handled by fitting with a student's t distribution. The system's derived parameters using radial velocities from both of the new approaches are provided in 
Table~\ref{tab:planet_params_final} (columns 1 and 2).

The precision reached thanks to method 1 and 2 means we have to take the circumbinary planet into account in order to properly compare them between one another and to TODCOR/TODMOR. As a nested sampler {\sc kima} can fit for the number of orbiting objects in a system (in our case binary and planet), and naturally marginalises over all parameters, including the number of orbiting bodies and their possible orbits. We discuss the planet's parameters in section~\ref{sec:planet}.

\subsection{Comparison of radial velocities}

After removing all Keplerian signals, we find that the residuals' root mean square (RMS) scatter of method 1 (SD-GP) is $39.9~\mathrm{m\,s^{-1}}$ for the primary star and $50.9~\mathrm{m\,s^{-1}}$ for the secondary star, while the RMS scatter of method 2 (CCF-GP) is $48.8 ~\mathrm{m\,s^{-1}}$ for the primary star, and $72.2~\mathrm{m\,s^{-1}}$ for the secondary star.

It is worth mentioning the root mean square (rms) values achieved by our new approaches are larger compared to the current state-of-the-art, a scatter of $10-15~\rm m,s^{-1}$ reported by \cite{Konacki_2009}. However, it is crucial to recognise that this increased scatter is likely inherent to the characteristics of the star itself. We have tested our methods on other double-lined binary systems, where we find that the scatter can reach down to photon noise.

Individual measurement uncertainties for the primary and secondary stars, measured by each method, range between $5.8-7.5~\mathrm{m\,s^{-1}}$ for method 1 and $4.7-13~\mathrm{m\,s^{-1}}$ for method 2. In Figure \ref{fig:comp2}, we plot radial velocities measured using method 1 against radial velocities measured using method 2. We find the mean absolute difference between the two approaches to be $26.9$ and $29.2~\mathrm{m\,s^{-1}}$ for the primary and secondary stars, respectively. These mean differences are lower than the measured scatter, but exceed the uncertainties estimated by the GP fits, which suggests the presence of a systematic bias between them\footnote{we tested method 1 and 2 on bright double-lined binaries from \citep{Konacki_2009,Konacki_2010}, observed with SOPHIE, and achieved accuracies of order $2-4~\rm m\,s^{-1}$, which will be the object of a follow-up paper.}. This bias could be due to various factors, such as the differences in the templates used for cross-correlation in method 2. In addition, method 1 might be more susceptible to the effects of stellar activity, which affect the accuracy of the radial velocities. Further analysis may be necessary to fully understand and quantify the sources of the observed differences.

In Figure~\ref{fig:compmod}, we show the radial velocity time series for each method, along with the binary+planet Keplerian models applied to them. We find radial velocities measured by our approaches are consistent with those measured by TODCOR and TODMOR within uncertainties. This suggests that our approaches are able to accurately measure the radial velocities of the primary and secondary stars. 
The distribution of residuals (Observed $-$ Calculated; O-C) for TODCOR, TODMOR, method 1, and method 2 is presented in Figure \ref{fig:compallrms} using violin plots. Each violin plot represents the distribution of velocities clustered around mean O-C values in $\rm m\,s^{-1}$, and the width of each plot functions like a histogram. Our proposed methods outperforms TODCOR and TODMOR in terms of root-mean-square (RMS) scatter (Figure \ref{fig:compallrms}), producing an improvement of a factor $\sim4$ and $\sim2$ respectively. This indicates the effectiveness of our new approaches in measuring double-lined binary radial velocities more precisely than before. Since both our methods agree between one another, we are also confident our measurements gained in precision without compromising in accuracy.

\subsection{The circumbinary planet within the TIC~172900988 system}\label{sec:planet}

The circumbinary planet is naturally detected and fitted by the Nested Sampler, however, we first describe a more frequentist approach as it might be closer to methods used in the binary star literature.

\begin{figure}
\includegraphics[width=0.495\textwidth]{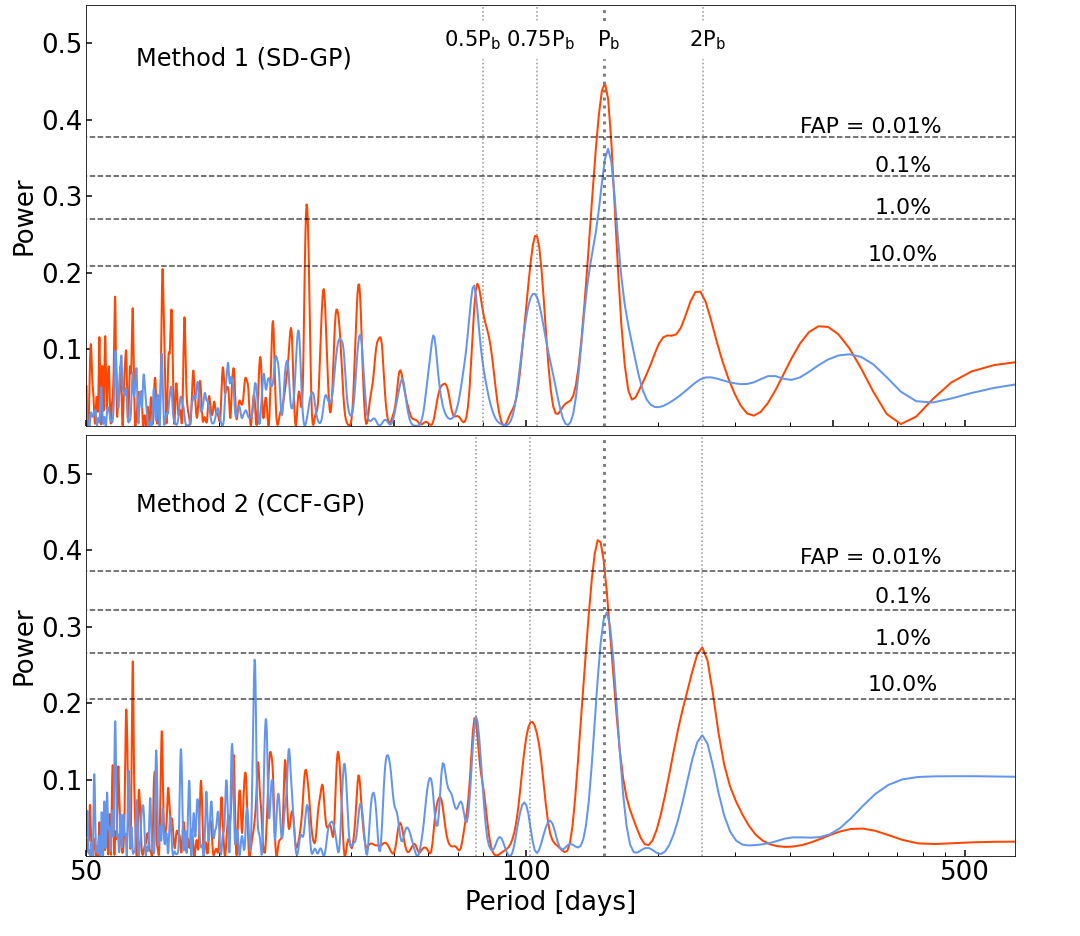}
\caption{Lomb-Scargle periodogram of TIC172900988 radial velocities for method 1 (top) and method 2 (bottom). The radial velocities for the primary (red) and the secondary (blue) are plotted after removing the binary motion. The three horizontal dashed lines indicate 10\%, 1\% and 0.1\% false alarm probabilities. The vertical dotted lines indicates the highly significant peak around 150 days and its harmonics. }
\label{fig:periodogram}
\end{figure}

We initially compute a generalised Lomb-Scargle (GLS) periodogram \citep{zechmeister_2009} for the radial velocities measured from the primary and the secondary stars, after having removed the best fitting Keplerian motion for the binary star. In Figure \ref{fig:periodogram} we display the resulting periodogram for method 1 SD-GP (top panel) and method 2 CCF-GP (bottom panel). Using 10\,000 bootstrap randomisation of the input data, we compute the false alarm probability (FAP) levels of 10, 1, 0.1 and 0.01\%. This calculation can be done independently for the primary and secondary radial velocities. The periodogram for radial velocities using both method 1 and method 2  show excess power at $P_{\rm pl}\sim151~\rm d$ with a ${\rm FAP} = 0.005\%$. After subtracting the signal $P_{\rm pl}$, the periodogram has no significant peak (Figure \ref{fig:res_period}).

\begin{figure}
\includegraphics[width=0.49\textwidth]{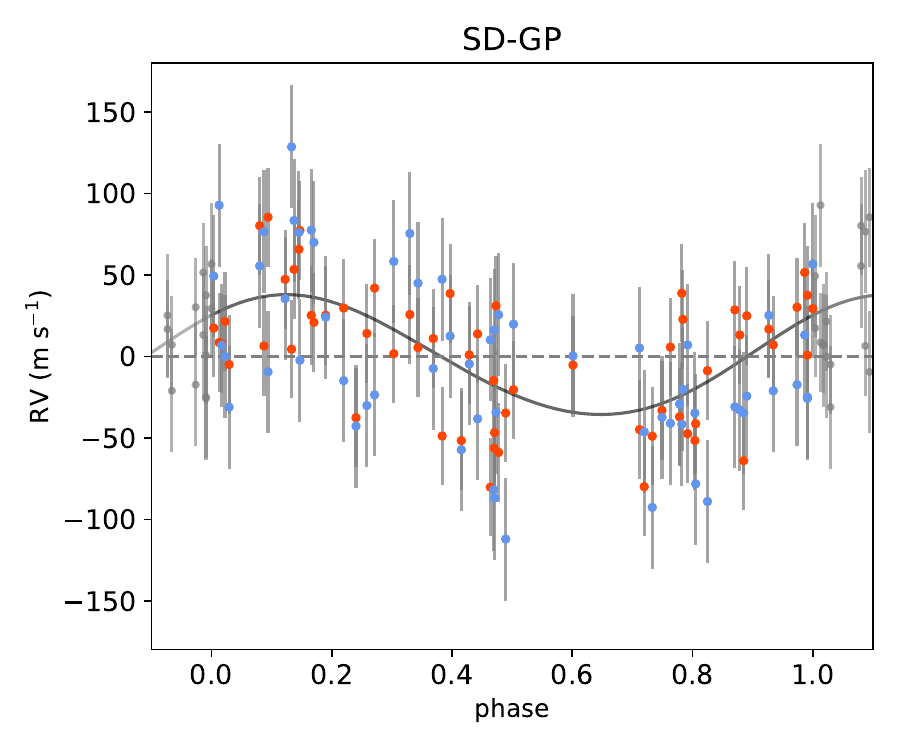}
\includegraphics[width=0.49\textwidth]{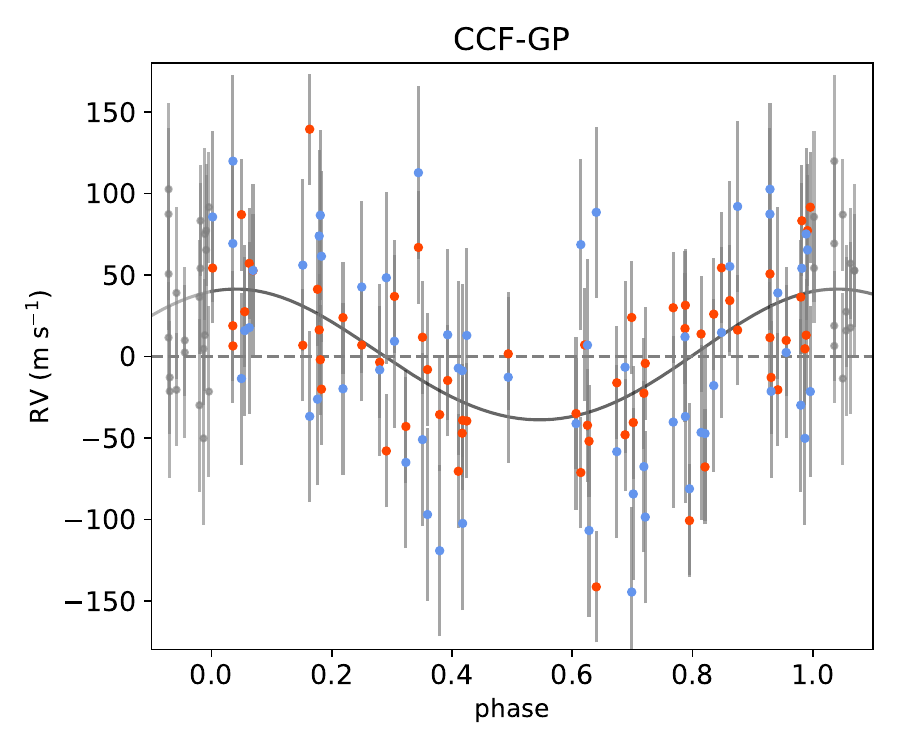}

\caption{
The residual RVs phase folded with corresponding best fit Keplerian circumbinary model (Table\ref{tab:planet_params_full}). The top and bottom panels correspond to our method 1 (SD-GP) and method 2 (SD-GP), respectively.}
\label{fig:rvfit}
\end{figure}

We perform a more thorough analysis of the data using the {\tt kima} analysis package which uses diffusive nested sampling \citep{faria_2018,Baycroft_2023}. {\tt kima} allows for Bayesian model comparison by computing the Bayes factor between a model with a binary and one planet to one with a binary but no planet from posterior samples generated by the algorithm. Using the Jeffrey's scale \citep{KassandRaftery_1995}, a Bayes factor ($\rm BF$) over 150 is considered strong evidence in favour of the more complex model (here binary+planet). Therefore we use this value as our confident detection threshold.
In Figure \ref{fig:rvfit}, we show the phased radial velocity data with the best-fit Keplerian model for the circumbinary planet (the binary having been removed). This is done for the data from method 1 (top panel) and method 2 (bottom panel).

The version of {\tt kima} we use fits for all the orbital elements of the binary, except $\Omega$ and $i$, but $i$ is known from the eclipsing geometry \citep{Kostov2021}. A different systemic velocity parameter is fit for each of the two stars. Keplerian models of the planet also include all orbital parameters except $\Omega$ and $i$. Two jitter terms are also fit by {\tt kima}, one for the primary and one for the secondary. To include outliers properly in our fit, we use Student's t statistics.

{\tt kima}'s fit of the SOPHIE data obtained on TIC~172900988 yields ${\rm BF} =  2\,300\,000$ using the SD-GP method (method 1) and ${\rm BF} = 16\,000$ using the CCF-GP method (method 2). Both approaches exceed the detection threshold and imply a confident detection of a circumbinary planet. The parameters for the planet (as well as the binary) are shown in 
Table \ref{tab:planet_params_final}. Since TIC~172900988 is a double-lined eclipsing binary we obtain the absolute mass of each stellar component at high precision. Since the planet's orbital plane at the time of the observations is close to perpendicular to the line of sight \citep{Kostov2021}, we measure a mass that could be considered an absolute mass as well. However, it is likely the planetary orbital plane inclination has precessed, and might be out of transitability \citep[e.g.][]{Martin2014}.
We use the median of posteriors from $\tt kima$ and their $1\sigma$ confidence region to produce our fit's parameters and uncertainties. We find all binary parameters to be statistically consistent with the analysis of \citep{Kostov2021}, with a few caveats. The binary period ($P_{\rm B}$) we find is inconsistent with any of the 6 solutions proposed by \citep{Kostov2021}, but it does lie within the range that these solutions cover\footnote{These six solutions consist of osculating elements and the binary periods are internally inconsistent with each other}. The argument of periastron ($\omega_{\rm B}$) that we measure is not consistent, at first glance. However, our measurements were taken some time after the \citet{Kostov2021} paper. If we correct for the apsidal precession of the binary orbit, our value of $\omega_{\rm B}$ is consistent with \citet{Kostov2021}. The value we get for the apsidal precession rate ($\dot{\omega}_{\rm B}$) is also consistent with the value quoted in \citet{Kostov2021}. This precession rate exceeds that expected from General Relativity, and is attributed to the third-body perturbations produced by the planet. Figure~\ref{fig:prec} shows the area of parameter-space a third body needs to have to produce and apsidal precession rate consistent with the observations. We overplot the location of the planetary parameters presented in this work and the solutions proposed by \citet{Kostov2021}. All solutions can reproduce the detected precession rate well. 

For the planet, we find $P_{\rm pl} \approx 150~\rm d$, and a mass $m_{\rm pl} \approx 2 ~\rm M_{\rm jup}$ ($\approx600~\rm M_\oplus$). While these are inconsistent with any of the six solutions proposed by \citet{Kostov2021}, we fit the planet with a Keplerian model and report mean parameters, where \citet{Kostov2021} fit with a dynamical model and report osculating parameters. In such a dynamically complex orbit, it is difficult to compare these parameters properly. We note an additional important caveat here: We fit a static Keplerian to the planetary orbit and obtain a mean orbital period. Other parameters such as the semi-major axis and the mass are then calculated using Kepler's law. However, due to the proximity of this orbit to the binary it is expected that non-Keplerian effects (such as apsidal precession of the planetary orbit) are present and the orbit would not conform to Kepler's law. Hence it is possible that the planet's true orbital distance and true mass are slightly larger than stated here. 

Orbital parameters between both methods (SD-GP and CCF-GP) are internally consistent and are presented in Table~\ref{tab:planet_params_final}, columns 1 and 2. Any posterior samples where the proposed planet crosses into the instability region (calculated using the formula in \citealt{holman_1999}) are excluded, as described in \citet{standing_2022}. We adopt the parameters in column 1 of Table \ref{tab:planet_params_final} as our preferred solution. We chose this solution as the SD-GP method results in a higher Bayes Factor than the CCF-GP. The method used for the parameters reported in columns 3 and 4 is described in section \ref{sec:12p}. Table \ref{tab:planet_params_full} shows the parameters obtained from the posterior sample with the "unstable" solutions left in. These are therefore the parameters simply consistent with the data without dynamical stability being considered.

\begin{figure}
\includegraphics[width=0.5\textwidth]{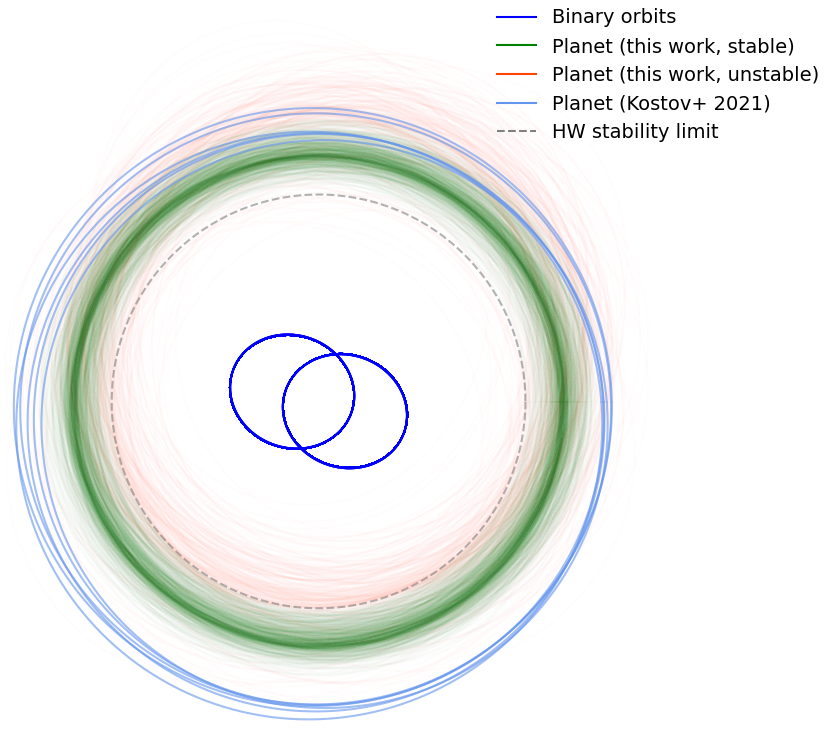}
\caption{Orbital configuration of TIC~172900988 showing the orbits of the binary and the planet. The red orbits are a random sample of 1000 posteriors from {\tt kima} fitting the radial velocities from SD-GP. The green orbits are the 6 suggested solutions from \citet{Kostov2021}, the dashed grey line is the stability limit as calculated by \citet{holman_1999}. The radial velocity data alone are fit (not including the 1-2 punch transit data).}
\label{fig:orbit_view_spec}
\end{figure}

Figure \ref{fig:orbit_view_spec} shows the orbital configuration of TIC~172900988, displaying the orbits of the binary and the planet. A sub-sample of 1000 posterior samples are drawn: if a sample crosses into the instability zone it gets shown in red, otherwise in green. The parameters in Table \ref{tab:planet_params_final} and the distributions shown in Figure \ref{fig:corner_12punch} correspond therefore to the orbits shown in green. The 6 solutions from \citet{Kostov2021} are also shown for comparison \footnote{These may not be a full representation of the solutions from \citet{Kostov2021} since they quote osculating elements and we plot them as if they were mean elements}.

The \texttt{kima} algorithm can generate detection limits for any further signals, following the method presented in \citet{standing_2022}: first all planetary detected signals are removed from the data (but the binary's orbital signature is kept), then \texttt{kima} is run once more and forced to fit a planetary signal (when presumably there are none left in the data). The resulting set of posterior samples corresponds to all signals that are compatible with the data, but have no statistically detectable signals. This method is an alternative to injection-recovery tests \citep[e.g.][]{Konacki_2009,Konacki_2010, martin_2019}, that allows to compute a detection limit efficiently over a large parameter space, while marginalising over all orbital elements. The detection limits for TIC~172900988 are shown in Figure \ref{fig:detlim_noplanet} and reveal that the SOPHIE data analysed using our two new methods produce very similar results and that those are sensitive to planets with masses of order Jupiter's out to periods as large as $1000~\rm d$ except for orbital periods around the one-year alias.

\begin{figure}
\includegraphics[width=0.5\textwidth]{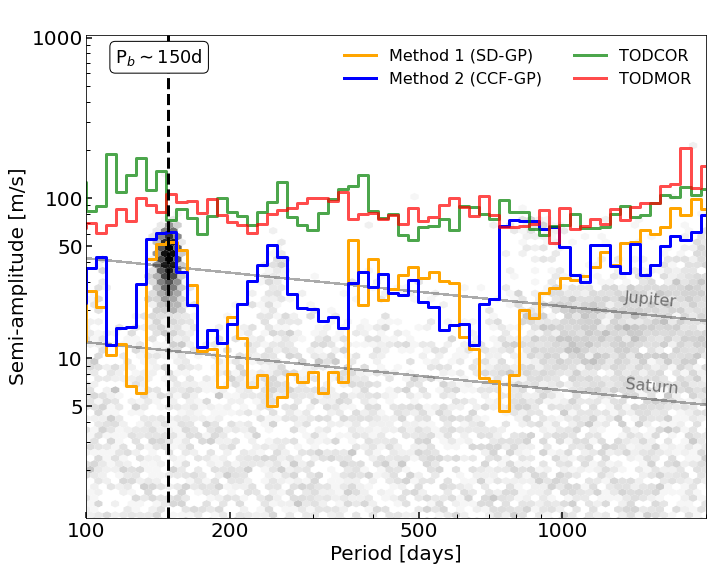}
\caption{Detection sensitivity to planets plotted as semi-amplitude as a function of orbital period of planets. The density of posterior samples are depicted as grey hexagonal bins. The solid green, red, orange and blue lines shows the detection limit from posterior samples for TODCOR, TODMOR, method 1 and method 2, respectively. The diagonal lines are anticipated signals of Saturn and Jupiter mass planet.  }
\label{fig:detlim_many}
\end{figure}

Finally, we run the same analysis on the TODCOR and TODMOR-produced radial velocities. These produced ${\rm BF} = 0.8$ and $0.7$ respectively, well below the accepted detection threshold, demonstrating that our new approaches out-performed TODCOR and TODMOR. We show a comparison of all the resulting detection limits in Figure \ref{fig:detlim_many}. The detection limits here are a little different than in Fig.~\ref{fig:detlim_noplanet}, since the planet is not formally detected with all approaches. To allow for a proper comparison between the detection limits generated using the different methods we do not remove the planetary signal when calculating the detection limits on data from the SD-GP and CCF-GP methods. To ensure that the parameter space was well-covered in these cases we then force {\tt kima} to fit 2 signals instead of the usual 1 signal.

\begin{table*}
    \centering
        \caption{Best-fit parameters for each methods both including and not including the 1-2 punch into the fit. Solutions that cross the instability limit \citep{holman_1999} are excluded here (parameters from the full posteriors can be found in Table \ref{tab:planet_params_full}). The parameters are determined along with their corresponding 1$\sigma$ uncertainties.}
    \label{tab:planet_params_final}
    \begin{tabular}{l|c|c|c|c}
        \hline
        &\bf{ADOPTED}&&&\\
        Parameters & Method 1 & Method 2 & Method 1 (with 1-2 punch) & Method 2 (with 1-2 punch) \\
         &SD-GP& CCF-GP&SD-GP& CCF-GP\\
        \hline
        Binary parameters&\\
        $P_{\rm B}$ [d] & $19.657878^{+0.000029}_{-0.000034}$ & $ 19.657861\pm0.000040 $ & $19.657874^{+0.000030}_{-0.000037}$ & $ 19.657869^{+0.000048}_{-0.000051} $\\
        $e_{\rm B}$  & $0.448234^{+0.000090}_{-0.000071}$& $0.44823\pm0.00010$ & $0.448230^{+0.000081}_{-0.000083}$& $0.44819\pm0.00013$\\
        $\omega_{\rm B}$ [rad] &$1.23147^{+0.00019}_{-0.00027}$& $1.23136^{+0.00035}_{-0.00038}$ &$1.23142^{+0.00022}_{-0.00024}$& $1.23112^{+0.00049}_{-0.00036}$\\
        $K_{\rm B}$ [$\mathrm{km~s^{-1}}$] & $58.5494^{+0.0079}_{-0.0054}$& $58.5279^{+0.0076}_{-0.0074}$ & $58.580\pm0.011$& $58.5428^{+0.0104}_{-0.0092}$\\ 
        $q_{\rm B}$  & $0.97196^{+0.00016}_{-0.00022}$& $0.97106^{+0.00025}_{-0.00023}$ & $0.97211\pm0.00022$& $0.97121^{+0.00029}_{-0.00027}$\\
        $\dot{\omega}_{\rm B} \rm[arcsec/yr]$  & $215^{+56}_{-59}$& $279^{+72}_{-70}$ & $229^{+62}_{-71}$ & $250^{+120}_{-110}$\\
        $T_{0,{\rm B}}$ [BJD] & $2\,459\,566.00250^{+0.00044}_{-0.00061}$ & $2\,459\,566.00210^{+0.00079}_{-0.00088}$ & $2\,459\,566.00267^{+0.00053}_{-0.00062}$ & $2\,459\,507.02830^{+0.00113}_{-0.00094}$\\
        
        \hline
        Planet parameters&&\\
        $P_{\rm pl}$ [d] & $151.2\pm1.8$ & $149.3\pm2.2 $ & $151.4\pm1.7$ & $149.9^{+2.4}_{-2.2} $ \\
        $e_{\rm pl}$ & $<0.11$ & $<0.11$ & $ 0.1243^{+0.0183}_{-0.0090} $ & $ 0.1251^{+0.0124}_{-0.0083} $\\
        $\omega_{\rm pl}$ [rad] & $6.0\pm1.8$ & $4.9\pm2.0$ & $5.00^{+0.31}_{-0.49}$ & $4.76\pm0.41$ \\
        $K_{\rm pl}$ [$\mathrm{m~s^{-1}}$] & $40.1^{+5.1}_{-5.3}$ & $44.0\pm6.8$ & $39.1^{+4.5}_{-4.3}$ & $40.5^{+6.3}_{-6.6}$\\ 
        $T_{0,{\rm pl}}$ [BJD] & $2\,459\,442^{+40}_{-45}$ & $2\,459\,425^{+45}_{-49}$ & $2\,459\,419.7^{+7.8}_{-12.0}$ & $2\,459\,421\pm10$ \\
        \hline
        Derived parameters&&\\
        $M_{\rm A}$ [$\mathrm{M_{\odot}}$] & $1.23681^{+0.00037}_{-0.00039}$& $1.23765^{+0.00051}_{-0.00046}$ & $1.23670\pm0.00036$& $1.23831^{+0.00060}_{-0.00063}$\\
        $M_{\rm B}$ [$\mathrm{M_{\odot}}$] & $1.20207^{+0.00033}_{-0.00026}$& $1.20184^{+0.00035}_{-0.00031}$ & $1.20221^{+0.00032}_{-0.00028}$& $1.20264^{+0.00048}_{-0.00046}$\\
        $m_{\rm pl}$ [$\mathrm{M_{Jup}}$] & $1.90\pm0.25$  & $2.07\pm0.32$ & $1.84\pm0.21$  & $2.09\pm0.27$\\
        $a_{\mathrm{bin}}$ [$\mathrm{AU}$] & $0.191879^{+0.000016}_{-0.000018}$& $0.191894\pm0.000021$ & $0.191879\pm0.000016$& $0.191931^{+0.000026}_{-0.000028}$\\
        $a_{\mathrm{pl}}$ [$\mathrm{AU}$]  & $0.7474^{+0.0057}_{-0.0061}$ & $0.7410^{+0.0069}_{-0.0075}$ & $0.7479^{+0.0051}_{-0.0056}$ & $0.7433^{+0.0079}_{-0.0072} $\\
        \hline
        Other parameters&&\\
        $V_{\mathrm{sys,A}}$ [$\mathrm{km~s^{-1}}$]  & $25.9865^{+0.0045}_{-0.0047}$ & $26.0252\pm0.0057$ & $26.0144\pm0.0045$ & $26.0573\pm0.0067$\\ 
        $V_{\mathrm{sys,B}}$ [$\mathrm{km~s^{-1}}$] & $26.0456^{+0.0062}_{-0.0060}$  & $26.0969^{+0.0087}_{-0.0089}$ & $26.0636^{+0.0059}_{-0.0061}$  & $26.116\pm0.010$\\ 
        \hline
    \end{tabular}
\end{table*}

\subsection{Including the 1-2 punch technique}\label{sec:12p}
The detection of the planet, and the posteriors of its orbital and physical parameters can be improved by combining the radial velocity data with some aspects of the transit data. Ultimately, a full photodynamical analysis would need to be performed, but this is beyond the scope of our paper. 

TIC~172900988 was discovered using the "1-2 punch" method \citep{Kostov2021}. Two transits within the same conjunction give an estimate of the planet's orbital period. The distance the planet has moved can be calculated from the position in its orbit of the transited star, at the time of each transit. The time between the transits and the distance travelled then allow us to calculate an estimate of the planet's orbital period (as in \citealt{Kostov2020}):
\begin{equation}
    \label{eq:12period}
    P_{\rm pl} = \frac{2\pi G M_{\rm bin}}{v^3} \left(\frac{e\sin\omega + \sin(\phi + \omega)}{\sqrt{1-e^2}}\right)^3,
\end{equation}
where $M_{\rm bin}$ is the total mass of the binary, $e$ and $\omega$ the eccentricity and argument of periastron of the planet, $\phi$ is the true anomaly at the point in the orbit that we are measuring and $v$ the average velocity in the plane of the sky with which the planet moved between both transit mid-points. Since we know the planet is at conjunction, we can use $\sin(\phi + \omega) \approx 1$\footnote{We note a difference from \citet{Kostov2020} where the value of this is quoted as -1, \citet{kostov_2016} use a value of 1, this may be down to the choice of reference frame}.

We alter our version of \texttt{kima} and add an extra feature, to include the "1-2 punch" information as part of the the sampling process. When a solution is proposed by the sampler, the predicted period from the "1-2 punch" is calculated and compared to the proposed period. This is then included in the likelihood calculation of the sample in the same way as an extra data point would be, assuming a Gaussian distribution. Our new log-likelihood is therefore:
\begin{equation}
    \log(\mathcal{L}) = \log(\mathcal{L}_{\rm RV}) -\frac{1}{2}\log(2\pi \sigma_{P_{12}}^2) - \frac{(P_{12} - P_{\rm pl})^2}{2 \sigma_{P_{12}}^2},
\end{equation}
where $\log(\mathcal{L}_{\rm RV})$ is the log-likelihood from the radial velocity data, $P_{\rm pl}$ is the period for the planet proposed as part of the sampling process, $P_{12}$ is the orbital period calculated using Eq.~\ref{eq:12period} using all other parameters proposed in the sample (e.g. $M_{\rm bin}, e$, etc). Finally,  $\sigma_{P_{12}}^2$ is the variance of $P_{12}$, which is derived from the uncertainty in the transit times propagated through Eq.~\ref{eq:12period}.

We run the analysis on TIC~172900988 again, with the extra input of two transits with mid-times at $2\,458\,883.390879\pm0.006188$ and $2\,458\,888.309427\pm0.003904$\footnote{The transit mid-times were determined using the Eclipsing Light Curve (ELC) code. 
The segment of the data containing a single event was isolated, 
and a model was fitted to the transit (or eclipse) profile. 
Then we execute the DE-MCMC code, and the median along with 1-sigma uncertainties 
of the posterior sample was considered as the best-fitting transit time.}. We fix the number of planets searched for in \texttt{kima} to 1. The parameters for the planet and binary obtained are shown in Table \ref{tab:planet_params_final} (columns 3 and 4). As with the previous analysis, any posterior samples where the proposed planet crosses into the instability are excluded.

A Keplerian fit of the radial velocity data probes the average parameters of the orbit over the time baseline, notably the average orbital period. A circumbinary planet, especially one like TIC~172900988\,b which is quite close to the inner binary will see its orbital parameters vary throughout its orbital path meaning that when parameters of the orbit are measured over a short time frame, they may not be representative of the average orbit. The 1-2 punch, method calculates the velocity and therefore the orbital period, over a short time frame. Using this method to constrain the average period might bring in a poorly understood bias. We therefore present the results from the combined radial velocity and 1-2 punch fit out of interest, but do not adopt these parameters as our preferred solution.

\begin{figure*}
\includegraphics[width=0.8\textwidth]{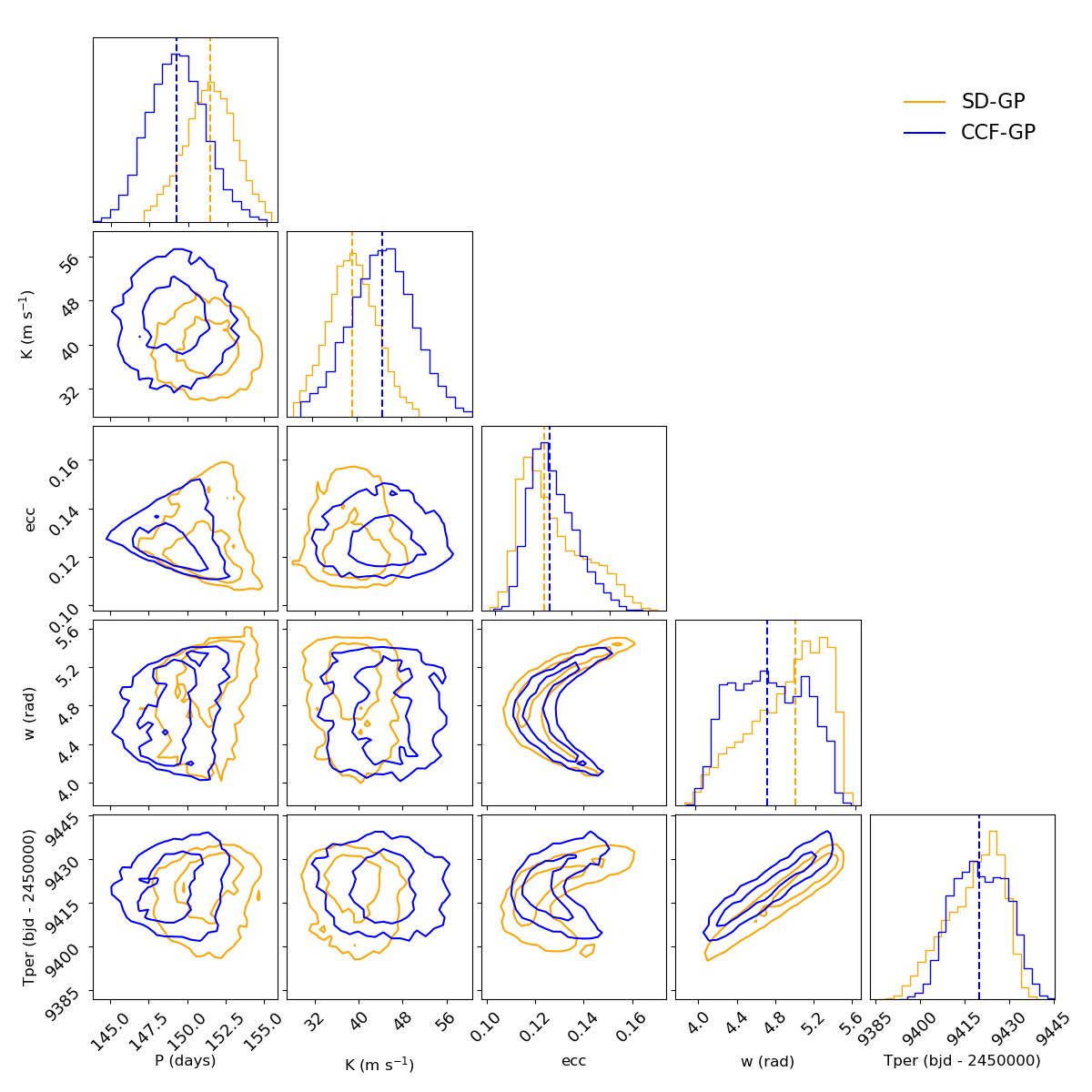}
\caption{Corner plot showing the distributions and correlations of the planetary parameters from the simultaneous fit of the radial velocity data and the 1-2 punch transit times. The contours are the 50th and 90th percentiles. Orange shows the results using the the radial velocity data using SD-GP, and blue from the CCF-GP.}
\label{fig:corner_12punch}
\end{figure*}

The posterior samples for the planet parameters are shown as a corner plot \citep{foreman-mackey_cornerpy_2016} in Figure \ref{fig:corner_12punch} for both the CCF-GP method and the SD-GP method. We note that the crescent-shaped correlations involving the eccentricity are expected. We also note that while we report the median and 1-$\sigma$ in Table \ref{tab:planet_params_final}, some of the parameters have non-Gaussian distributions (in particular the eccentricity ($e_{\rm pl}$) and argument of periastron ($\omega_{\rm pl}$)).

We find these new solutions are consistent with solutions of fitting just the radial velocities or just the average orbital velocity at conjunction between the transits. The combined fit suggests the circumbinary planet's orbit must have an eccentricity $e_{\rm pl}>0.1$ and an argument of periastron $4.35 \leq \omega_{\rm pl} \leq5.31 ~\rm rad$.

Figure \ref{fig:orbit_view_12_punch} shows the orbital configuration of TIC~172900988. This is the counterpart to Figure \ref{fig:orbit_view_spec} generated from posterior samples obtained from the {\tt kima} analysis which included the 1-2 punch and using the SD-GP radial velocities.

\section{Conclusions}\label{sec:conclusion}

In this work\, we focus on the development of two new data-driven approaches to accurately measure radial velocities in double-lined binary systems. Despite being brighter and more precise in principle, the time-varying blending of the two stars' spectral lines makes accurate radial velocity measurement challenging. Previous methods \citep{Konacki_2009} been shown to have a typical scatter of $10-15~\rm m\,s^{-1}$ that prevents the detection of most orbiting circumbinary planets. 

In this paper, we introduce two new methods based on 
GP regression inspired by \citet{Czekala_2017}. The first method applies the GP in the spectral domain and the second is applied on cross-correlated spectra. We compare the precision and accuracy of our radial-velocity to two widely used methods: TODCOR and TODMOR (\citealt{Mazeh_1994, zucker_2004}).

To perform the comparison, we analyse 62 SOPHIE spectra of the binary TIC~172900988, a binary system which was also proposed to host a circumbinary planet \citep{Kostov2021}. We show that our two methods outperform both TODCOR and TODMOR, neither of which could recover the planet whereas both our GP approaches successfully detect a circumbinary planet. However, its parameters are found statistically different from previously published solutions \citep{Kostov2021}. TIC~172900988\,b will now integrate the BEBOP catalogue for circumbinary exoplanets detected with radial-velocities as its second entry.

The RMS achieved by our new approaches are $\approx 50$ and $70~ \rm m\,s^{-1}$ for the primary and secondary stars of TIC~172900988, respectively, are larger than our measurement uncertainties, and larger than the typical scatter reported for double-lined binaries in \citet[$10-15~\rm m\,s^{-1}$;][]{Konacki_2009,Konacki_2010}. We speculate this increased scatter is most likely of stellar origin. Both stellar components are fairly high-mass stars. We note that TODCOR and TODMOR, both recognised methods of radial velocity extraction, also produce a high scatter. In those cases, TODCOR and TODMOR remain unable to detect the planetary signal which has a semi-amplitude $K_{\rm pl} \sim 42 ~\rm m\,s^{-1}$. The fact our approaches both manage to overcome some of that scatter emphasises the limitations of existing techniques when dealing with systems characterised by substantial scatter. The detection of a circumbinary planet in TIC~172900988 showcases the effectiveness of our data-driven methods in uncovering planetary signals even in challenging double-lined binary systems. We highlight here that should a circumbinary planet similar to the parameters of TIC~172900988\,b have been present in a quieter binary star system, traditional methods such as TODCOR, TODMOR and the tomographic disentangling method have the nominal accuracy to detect it.

Our two new methods are a step forward, but there is always room for improvement. 
Further refinements and optimisations to these methods may lead to even more accurate and precise radial velocity measurements, particularly with the spectral decomposition.

Firstly, we recognise that our analysis does not account for possible contamination in the radial velocities obtained from each chunk of the spectra. Specifically, we do not consider the effects of stellar activity on our results. Also the chunking can be improved to avoid areas that are poor in absorption lines, and avoid areas that include band known to be highly variable such as $\rm H\alpha$. In order to improve the accuracy and precision of our measurements, we plan to develop a more sophisticated approach that can identify regions of the spectra that are affected by these factors. 

Secondly, we recognise that there is still much to be learned about the astrophysical properties of the binary stars themselves. In the case of TIC~172900988, both stars are equal mass. The CCF and spectral decomposition methods might need to be adapted to non-equal mass binaries to account for the their differing spectral types. 

In addition, we expect that the spectral deconvolution method will yield accurate spectra for both stars individually when all wavelength chunks are combined together. Such a spectrum could be used to constrain their properties such as their temperature, $v\sin i$ and metallicity, an important parameter to relate planet presence to planet formation \citep[e.g.][]{Santos_2004, Adibekyan_2013}.

Our methods open the door to extend the search for circumbinary planets using the radial-velocity method beyond single-lined binaries. With our two new approaches, it is highly probable that the discovery of circumbinary planets will be enhanced in the future.  Finally, we highlight the success of our two new methods in being the first to detect a circumbinary planet using radial-velocities in a double-lined binary. Importantly this detection is made independently of any other data. Interestingly our results produce planetary parameters different from those previously published demonstrating the need for radial-velocity follow-up of circumbinary planet candidates identified with the transit method.

\section*{Acknowledgements}

We thank Tsevi Mazeh for suggesting we compare our methods to TODCOR and TODMOR, during the 2022 EAS conference in Valencia where preliminary results were presented. 

This paper is based on observations collected the Observatoire de Haute Provence (OHP). Here, we would also like to acknowledge the staff, particularly the night assistants, 
at the OHP for their dedication and hard work, 
especially during the COVID pandemic. 

This research received funding from 
the European Research Council (ERC) under the European Union's Horizon 2020 
research and innovation programme (grant agreement n° 803193/BEBOP) 
and from the Leverhulme Trust (research project grant n° RPG-2018-418). 
The French group acknowledges financial support from the French Programme 
National de Planétologie (PNP, INSU). 
The acquisition of data was made possible by a series of allocations through the French PNP. 
The computations described in this paper were performed using the University of 
Birmingham's BlueBEAR HPC service (available at http://www.birmingham.ac.uk/bear).

\section*{Data Availability}

The radial velocity data are included in the appendix of this paper. The reduced spectra are available at the SOPHIE archive: \href{http://atlas.obs-hp.fr/sophie/}{http://atlas.obs-hp.fr/sophie/}.


\bibliographystyle{mnras}
\bibliography{main} 

\begin{thebibliography}{}
\makeatletter
\relax
\def\mn@urlcharsother{\let\do\@makeother \do\$\do\&\do\#\do\^\do\_\do\%\do\~}
\def\mn@doi{\begingroup\mn@urlcharsother \@ifnextchar [ {\mn@doi@} {\mn@doi@[]}}
\def\mn@doi@[#1]#2{\def\@tempa{#1}\ifx\@tempa\@empty \href {http://dx.doi.org/#2} {doi:#2}\else \href {http://dx.doi.org/#2} {#1}\fi \endgroup}
\def\mn@eprint#1#2{\mn@eprint@#1:#2::\@nil}
\def\mn@eprint@arXiv#1{\href {http://arxiv.org/abs/#1} {{\tt arXiv:#1}}}
\def\mn@eprint@dblp#1{\href {http://dblp.uni-trier.de/rec/bibtex/#1.xml} {dblp:#1}}
\def\mn@eprint@#1:#2:#3:#4\@nil{\def\@tempa {#1}\def\@tempb {#2}\def\@tempc {#3}\ifx \@tempc \@empty \let \@tempc \@tempb \let \@tempb \@tempa \fi \ifx \@tempb \@empty \def\@tempb {arXiv}\fi \@ifundefined {mn@eprint@\@tempb}{\@tempb:\@tempc}{\expandafter \expandafter \csname mn@eprint@\@tempb\endcsname \expandafter{\@tempc}}}

\bibitem[\protect\citeauthoryear{{Adibekyan} et~al.,}{{Adibekyan} et~al.}{2013}]{Adibekyan_2013}
{Adibekyan} V.~Z.,  et~al., 2013, \mn@doi [\aap] {10.1051/0004-6361/201322551}, \href {https://ui.adsabs.harvard.edu/abs/2013A&A...560A..51A} {560, A51}

\bibitem[\protect\citeauthoryear{Aigrain, Parviainen  \& Pope}{Aigrain et~al.}{2016}]{aigrain_2016}
Aigrain S.,  Parviainen H.,   Pope B. J.~S.,  2016, \mn@doi [Monthly Notices of the Royal Astronomical Society] {10.1093/mnras/stw706}, 459, 2408

\bibitem[\protect\citeauthoryear{{Baranne} et~al.,}{{Baranne} et~al.}{1996}]{Baranne_1996}
{Baranne} et~al., 1996, \mn@doi [Astron. Astrophys. Suppl. Ser.] {10.1051/aas:1996251}, 119, 373

\bibitem[\protect\citeauthoryear{{Baycroft}, {Triaud}, {Faria}, {Correia}  \& {Standing}}{{Baycroft} et~al.}{2023}]{Baycroft_2023}
{Baycroft} T.~A.,  {Triaud} A. H.~M.~J.,  {Faria} J.,  {Correia} A. C.~M.,   {Standing} M.~R.,  2023, \mn@doi [\mnras] {10.1093/mnras/stad607}, \href {https://ui.adsabs.harvard.edu/abs/2023MNRAS.521.1871B} {521, 1871}

\bibitem[\protect\citeauthoryear{{Borucki} \& {Summers}}{{Borucki} \& {Summers}}{1984}]{Borucki_1984}
{Borucki} W.~J.,  {Summers} A.~L.,  1984, \mn@doi [\icarus] {10.1016/0019-1035(84)90102-7}, \href {https://ui.adsabs.harvard.edu/abs/1984Icar...58..121B} {58, 121}

\bibitem[\protect\citeauthoryear{{Bouchy} et~al.,}{{Bouchy} et~al.}{2009}]{Bouchy_2009}
{Bouchy} F.,  et~al., 2009, \mn@doi [\aap] {10.1051/0004-6361/200912427}, \href {https://ui.adsabs.harvard.edu/abs/2009A&A...505..853B} {505, 853}

\bibitem[\protect\citeauthoryear{Brewer \& Foreman-Mackey}{Brewer \& Foreman-Mackey}{2018}]{brewer_dnest4_2018}
Brewer B.~J.,  Foreman-Mackey D.,  2018, \mn@doi [Journal of Statistical Software] {10.18637/jss.v086.i07}, 86, 1

\bibitem[\protect\citeauthoryear{Byrd, Lu, Nocedal  \& Zhu}{Byrd et~al.}{1995}]{Byrd_1995}
Byrd R.~H.,  Lu P.,  Nocedal J.,   Zhu C.,  1995, \mn@doi [SIAM Journal on Scientific Computing] {10.1137/0916069}, 16, 1190

\bibitem[\protect\citeauthoryear{{Czekala}, {Mandel}, {Andrews}, {Dittmann}, {Ghosh}, {Montet}  \& {Newton}}{{Czekala} et~al.}{2017}]{Czekala_2017}
{Czekala} I.,  {Mandel} K.~S.,  {Andrews} S.~M.,  {Dittmann} J.~A.,  {Ghosh} S.~K.,  {Montet} B.~T.,   {Newton} E.~R.,  2017, \mn@doi [\apj] {10.3847/1538-4357/aa6aab}, \href {https://ui.adsabs.harvard.edu/abs/2017ApJ...840...49C} {840, 49}

\bibitem[\protect\citeauthoryear{{Doyle} et~al.,}{{Doyle} et~al.}{2011}]{Doyle_2011}
{Doyle} L.~R.,  et~al., 2011, \mn@doi [Science] {10.1126/science.1210923}, \href {https://ui.adsabs.harvard.edu/abs/2011Sci...333.1602D} {333, 1602}

\bibitem[\protect\citeauthoryear{{Faria}, {Santos}, {Figueira}  \& {Brewer}}{{Faria} et~al.}{2018}]{faria_2018}
{Faria} J.~P.,  {Santos} N.~C.,  {Figueira} P.,   {Brewer} B.~J.,  2018, \mn@doi [The Journal of Open Source Software] {10.21105/joss.00487}, \href {https://ui.adsabs.harvard.edu/abs/2018JOSS....3..487F} {3, 487}

\bibitem[\protect\citeauthoryear{Foreman-Mackey}{Foreman-Mackey}{2016}]{foreman-mackey_cornerpy_2016}
Foreman-Mackey D.,  2016, \mn@doi [The Journal of Open Source Software] {10.21105/joss.00024}, 1, 24

\bibitem[\protect\citeauthoryear{Foreman-Mackey, Hogg, Lang  \& Goodman}{Foreman-Mackey et~al.}{2013}]{Foreman-Mackey_2013}
Foreman-Mackey D.,  Hogg D.~W.,  Lang D.,   Goodman J.,  2013, \mn@doi [Publications of the Astronomical Society of the Pacific] {10.1086/670067}, 125, 306

\bibitem[\protect\citeauthoryear{{Foreman-Mackey}, {Agol}, {Ambikasaran}  \& {Angus}}{{Foreman-Mackey} et~al.}{2017}]{Foreman-Mackey_2017}
{Foreman-Mackey} D.,  {Agol} E.,  {Ambikasaran} S.,   {Angus} R.,  2017, \mn@doi [\aj] {10.3847/1538-3881/aa9332}, \href {https://ui.adsabs.harvard.edu/abs/2017AJ....154..220F} {154, 220}

\bibitem[\protect\citeauthoryear{Goodman \& Weare}{Goodman \& Weare}{2010}]{goodman2010ensemble}
Goodman J.,  Weare J.,  2010, Communications in applied mathematics and computational science, 5, 65

\bibitem[\protect\citeauthoryear{{Holman} \& {Wiegert}}{{Holman} \& {Wiegert}}{1999}]{holman_1999}
{Holman} M.~J.,  {Wiegert} P.~A.,  1999, \mn@doi [\aj] {10.1086/300695}, \href {https://ui.adsabs.harvard.edu/abs/1999AJ....117..621H} {117, 621}

\bibitem[\protect\citeauthoryear{{Husser}, {Wende-von Berg}, {Dreizler}, {Homeier}, {Reiners}, {Barman}  \& {Hauschildt}}{{Husser} et~al.}{2013}]{husser_2013}
{Husser} T.~O.,  {Wende-von Berg} S.,  {Dreizler} S.,  {Homeier} D.,  {Reiners} A.,  {Barman} T.,   {Hauschildt} P.~H.,  2013, \mn@doi [\aap] {10.1051/0004-6361/201219058}, \href {https://ui.adsabs.harvard.edu/abs/2013A&A...553A...6H} {553, A6}

\bibitem[\protect\citeauthoryear{Kass \& Raftery}{Kass \& Raftery}{1995}]{KassandRaftery_1995}
Kass R.~E.,  Raftery A.~E.,  1995, \mn@doi [Journal of the American Statistical Association] {10.1080/01621459.1995.10476572}, 90, 773

\bibitem[\protect\citeauthoryear{{Konacki}, {Muterspaugh}, {Kulkarni}  \& {He{\l}miniak}}{{Konacki} et~al.}{2009}]{Konacki_2009}
{Konacki} M.,  {Muterspaugh} M.~W.,  {Kulkarni} S.~R.,   {He{\l}miniak} K.~G.,  2009, \mn@doi [\apj] {10.1088/0004-637X/704/1/513}, \href {https://ui.adsabs.harvard.edu/abs/2009ApJ...704..513K} {704, 513}

\bibitem[\protect\citeauthoryear{{Konacki}, {Muterspaugh}, {Kulkarni}  \& {He{\l}miniak}}{{Konacki} et~al.}{2010}]{Konacki_2010}
{Konacki} M.,  {Muterspaugh} M.~W.,  {Kulkarni} S.~R.,   {He{\l}miniak} K.~G.,  2010, \mn@doi [\apj] {10.1088/0004-637X/719/2/1293}, \href {https://ui.adsabs.harvard.edu/abs/2010ApJ...719.1293K} {719, 1293}

\bibitem[\protect\citeauthoryear{{Kostov} et~al.,}{{Kostov} et~al.}{2016}]{kostov_2016}
{Kostov} V.~B.,  et~al., 2016, \mn@doi [\apj] {10.3847/0004-637X/827/1/86}, \href {https://ui.adsabs.harvard.edu/abs/2016ApJ...827...86K} {827, 86}

\bibitem[\protect\citeauthoryear{{Kostov} et~al.,}{{Kostov} et~al.}{2020}]{Kostov2020}
{Kostov} V.~B.,  et~al., 2020, \mn@doi [\aj] {10.3847/1538-3881/ab8a48}, \href {https://ui.adsabs.harvard.edu/abs/2020AJ....159..253K} {159, 253}

\bibitem[\protect\citeauthoryear{{Kostov} et~al.,}{{Kostov} et~al.}{2021}]{Kostov2021}
{Kostov} V.~B.,  et~al., 2021, \mn@doi [\aj] {10.3847/1538-3881/ac223a}, \href {https://ui.adsabs.harvard.edu/abs/2021AJ....162..234K} {162, 234}

\bibitem[\protect\citeauthoryear{{Kovaleva}, {Malkov}, {Yungelson}  \& {Chulkov}}{{Kovaleva} et~al.}{2016}]{Kovaleva_2016}
{Kovaleva} D.,  {Malkov} O.,  {Yungelson} L.,   {Chulkov} D.,  2016, \mn@doi [Baltic Astronomy] {10.1515/astro-2017-0261}, \href {https://ui.adsabs.harvard.edu/abs/2016BaltA..25..419K} {25, 419}

\bibitem[\protect\citeauthoryear{{Martin}}{{Martin}}{2018}]{Martin2018}
{Martin} D.~V.,  2018, {Populations of Planets in Multiple Star Systems}.
Springer International Publishing, \mn@doi{10.1007/978-3-319-55333-7_156}, \url {https://doi.org/10.1007/978-3-319-55333-7_156}

\bibitem[\protect\citeauthoryear{{Martin} \& {Triaud}}{{Martin} \& {Triaud}}{2014}]{Martin2014}
{Martin} D.~V.,  {Triaud} A. H.~M.~J.,  2014, \mn@doi [\aap] {10.1051/0004-6361/201323112}, \href {https://ui.adsabs.harvard.edu/abs/2014A&A...570A..91M} {570, A91}

\bibitem[\protect\citeauthoryear{{Martin} et~al.,}{{Martin} et~al.}{2019}]{martin_2019}
{Martin} D.~V.,  et~al., 2019, \mn@doi [\aap] {10.1051/0004-6361/201833669}, \href {https://ui.adsabs.harvard.edu/abs/2019A&A...624A..68M} {624, A68}

\bibitem[\protect\citeauthoryear{{Mayor} \& {Queloz}}{{Mayor} \& {Queloz}}{1995}]{Mayor1995}
{Mayor} M.,  {Queloz} D.,  1995, \mn@doi [\nat] {10.1038/378355a0}, \href {https://ui.adsabs.harvard.edu/abs/1995Natur.378..355M} {378, 355}

\bibitem[\protect\citeauthoryear{{Mazeh} \& {Zucker}}{{Mazeh} \& {Zucker}}{1994}]{Mazeh_1994}
{Mazeh} T.,  {Zucker} S.,  1994, \mn@doi [\apss] {10.1007/BF00984538}, \href {https://ui.adsabs.harvard.edu/abs/1994Ap&SS.212..349M} {212, 349}

\bibitem[\protect\citeauthoryear{{Pepe}, {Mayor}, {Galland}, {Naef}, {Queloz}, {Santos}, {Udry}  \& {Burnet}}{{Pepe} et~al.}{2002}]{Pepe_2002}
{Pepe} F.,  {Mayor} M.,  {Galland} F.,  {Naef} D.,  {Queloz} D.,  {Santos} N.~C.,  {Udry} S.,   {Burnet} M.,  2002, \mn@doi [\aap] {10.1051/0004-6361:20020433}, \href {https://ui.adsabs.harvard.edu/abs/2002A&A...388..632P} {388, 632}

\bibitem[\protect\citeauthoryear{{Perruchot} et~al.,}{{Perruchot} et~al.}{2008}]{Perruchot_2008}
{Perruchot} S.,  et~al., 2008, in {McLean} I.~S.,  {Casali} M.~M.,  eds,  Society of Photo-Optical Instrumentation Engineers (SPIE) Conference Series Vol. 7014, Ground-based and Airborne Instrumentation for Astronomy II. p. 70140J, \mn@doi{10.1117/12.787379}

\bibitem[\protect\citeauthoryear{{Rajpaul}, {Aigrain}  \& {Buchhave}}{{Rajpaul} et~al.}{2020}]{Rajpaul_2020}
{Rajpaul} V.~M.,  {Aigrain} S.,   {Buchhave} L.~A.,  2020, \mn@doi [\mnras] {10.1093/mnras/stz3599}, \href {https://ui.adsabs.harvard.edu/abs/2020MNRAS.492.3960R} {492, 3960}

\bibitem[\protect\citeauthoryear{{Rasmussen} \& {Williams}}{{Rasmussen} \& {Williams}}{2006}]{rasmussen_2006}
{Rasmussen} C.~E.,  {Williams} C. K.~I.,  2006, {Gaussian Processes for Machine Learning}.
MIT Press

\bibitem[\protect\citeauthoryear{{Santos}, {Israelian}  \& {Mayor}}{{Santos} et~al.}{2004}]{Santos_2004}
{Santos} N.~C.,  {Israelian} G.,   {Mayor} M.,  2004, \mn@doi [\aap] {10.1051/0004-6361:20034469}, \href {https://ui.adsabs.harvard.edu/abs/2004A&A...415.1153S} {415, 1153}

\bibitem[\protect\citeauthoryear{{Schneider}}{{Schneider}}{1994}]{Schneider_1994}
{Schneider} J.,  1994, \mn@doi [\planss] {10.1016/0032-0633(94)90075-2}, \href {https://ui.adsabs.harvard.edu/abs/1994P&SS...42..539S} {42, 539}

\bibitem[\protect\citeauthoryear{{Socia} et~al.,}{{Socia} et~al.}{2020}]{Socia2020}
{Socia} Q.~J.,  et~al., 2020, \mn@doi [\aj] {10.3847/1538-3881/ab665b}, \href {https://ui.adsabs.harvard.edu/abs/2020AJ....159...94S} {159, 94}

\bibitem[\protect\citeauthoryear{{Standing} et~al.,}{{Standing} et~al.}{2022}]{standing_2022}
{Standing} M.~R.,  et~al., 2022, \mn@doi [\mnras] {10.1093/mnras/stac113}, \href {https://ui.adsabs.harvard.edu/abs/2022MNRAS.511.3571S} {511, 3571}

\bibitem[\protect\citeauthoryear{Standing et~al.,}{Standing et~al.}{2023}]{Standing2023}
Standing M.~R.,  et~al., 2023, \mn@doi [Nature Astronomy] {10.1038/s41550-023-01948-4}, 7, 702

\bibitem[\protect\citeauthoryear{{Triaud} et~al.,}{{Triaud} et~al.}{2017}]{Triaud_2017}
{Triaud} A. H.~M.~J.,  et~al., 2017, \mn@doi [\aap] {10.1051/0004-6361/201730993}, \href {https://ui.adsabs.harvard.edu/abs/2017A&A...608A.129T} {608, A129}

\bibitem[\protect\citeauthoryear{Triaud et~al.,}{Triaud et~al.}{2022}]{triaud_2022}
Triaud A. H. M.~J.,  et~al., 2022, \mn@doi [Monthly Notices of the Royal Astronomical Society] {10.1093/mnras/stab3712}, 511, 3561

\bibitem[\protect\citeauthoryear{{Zechmeister} \& {K{\"u}rster}}{{Zechmeister} \& {K{\"u}rster}}{2009}]{zechmeister_2009}
{Zechmeister} M.,  {K{\"u}rster} M.,  2009, \mn@doi [\aap] {10.1051/0004-6361:200811296}, \href {https://ui.adsabs.harvard.edu/abs/2009A&A...496..577Z} {496, 577}

\bibitem[\protect\citeauthoryear{{Zucker} \& {Mazeh}}{{Zucker} \& {Mazeh}}{1994}]{zucker_1994}
{Zucker} S.,  {Mazeh} T.,  1994, \mn@doi [\apj] {10.1086/173605}, \href {https://ui.adsabs.harvard.edu/abs/1994ApJ...420..806Z} {420, 806}

\bibitem[\protect\citeauthoryear{{Zucker}, {Mazeh}, {Santos}, {Udry}  \& {Mayor}}{{Zucker} et~al.}{2004}]{zucker_2004}
{Zucker} S.,  {Mazeh} T.,  {Santos} N.~C.,  {Udry} S.,   {Mayor} M.,  2004, \mn@doi [\aap] {10.1051/0004-6361:20040384}, \href {https://ui.adsabs.harvard.edu/abs/2004A&A...426..695Z} {426, 695}

\makeatother
\end{thebibliography}


\appendix

\section{Some extra material}

\begin{table*}
    \centering
        \caption{Best-fit parameters for each methods both including and not including the 1-2 punch into the fit. Solutions that cross the instability limit \citep{holman_1999} are not excluded here so these parameters are from the full posterior (parameters from the posteriors excluding unstable samples can be found in Table \ref{tab:planet_params_final}) The parameters are determined along with their corresponding 1$\sigma$ uncertainties.}
    \label{tab:planet_params_full}
    \begin{tabular}{l|c|c|c|c}
        \hline
        Parameters & Method 1 & Method 2 & Method 1 (with 1-2 punch) & Method 2 (with 1-2 punch) \\
         &SD-GP& CCF-GP&SD-GP& CCF-GP\\
        \hline
        Binary parameters&\\
        $P_{\rm B}$ [d] & $19.657878^{+0.000030}_{-0.000035}$ & $ 19.657861\pm0.000039 $ & $19.657872^{+0.000033}_{-0.000038}$ & $ 19.657865^{+0.000037}_{-0.000049} $\\
        $e_{\rm B}$  & $0.448234^{+0.000090}_{-0.000071}$& $0.44823\pm0.00010$ & $0.448232\pm0.000083$& $0.44821\pm0.00011$\\
        $\omega_{\rm B}$ [rad] &$1.23146^{+0.00020}_{-0.00027}$& $1.23136^{+0.00034}_{-0.00038}$ &$1.23142^{+0.00022}_{-0.00026}$& $1.23132^{+0.00042}_{-0.00038}$\\
        $K_{\rm B}$ [$\mathrm{km~s^{-1}}$] & $58.5494^{+0.0079}_{-0.0054}$& $58.5278\pm0.0076$ & $58.5557^{+0.0067}_{-0.0075}$& $58.5320\pm0.0085$\\ 
        $q_{\rm B}$  & $0.97196^{+0.00016}_{-0.00022}$& $0.97106\pm0.00025$ & $0.97211^\pm0.00022$& $0.97118\pm0.00026$\\
        $\dot{\omega}_{\rm B} \rm[arcsec/yr]$  & $215\pm59$& $279^{+73}_{-69}$ & $ 230^{+62}_{-68} $ & $ 279^{+87}_{-77} $\\
        $T_{0,{\rm B}}$ [BJD] & $2\,459\,566.00248^{+0.00044}_{0.00060}$ & $2\,459\,566.00209^{+0.00080}_{-0.00087}$ & $2\,459\,566.00267^{+0.00053}_{-0.00067}$ & $2\,459\,566.00236^{+0.00089}_{-0.00086}$\\
        
        \hline
        Planet parameters&&\\
        $P_{\rm pl}$ [d] & $151.3\pm1.8$ & $149.2^{+2.2}_{-2.5} $ & $151.7^{+1.4}_{-1.7}$ & $148.6^{+2.1}_{-2.4} $ \\
        $e_{\rm pl}$ & $<0.21$ & $<0.19$ & $ 0.153^{+0.095}_{-0.034} $ & $ 0.141^{+0.059}_{-0.019} $\\
        $\omega_{\rm pl}$ [rad] & $6.0\pm1.5$ & $4.9^{+1.8}_{-2.0}$ & $5.37^{+0.34}_{-0.69}$ & $4.57^{+0.75}_{-0.61} $ \\
        $K_{\rm pl}$ [$\mathrm{m~s^{-1}}$] & $40.0\pm5.2$ & $43.8\pm6.9$ & $39.1\pm4.4$ & $44.6\pm5.9$\\ 
        $T_{0,{\rm pl}}$ [BJD] & $2\,459\,442\pm36$ & $2\,459\,425^{+43}_{-45}\pm36$ & $2\,459\,428.6^{+8.2}_{-16.4}$ & $2\,459\,417^{+17}_{-14}$ \\
        \hline
        Derived parameters&&\\
        $M_{\rm A}$ [$\mathrm{M_{\odot}}$] & $1.23681^{+0.00037}_{-0.00039}$& $1.23766^{+0.00051}_{-0.00047}$ & $1.23670^\pm0.00036$& $1.23763^{+0.00056}_{-0.00052}$\\
        $M_{\rm B}$ [$\mathrm{M_{\odot}}$] & $1.20207^{+0.00032}_{-0.00027}$& $1.20184^{+0.00035}_{-0.00031}$ & $1.20221^{+0.00032}_{-0.00029}$& $1.20196^{+0.00046}_{-0.00036}$\\
        $m_{\rm pl}$ [$\mathrm{M_{Jup}}$] & $1.88\pm0.25$  & $2.05\pm0.33$ & $1.82\pm0.21$  & $2.08\pm0.28$\\
        $a_{\mathrm{bin}}$ [$\mathrm{AU}$] & $0.191879^{+0.000016}_{-0.000018}$& $0.191894^{+0.000021}_{-0.000019}$ & $0.191879\pm0.000017$& $0.191896^{+0.000026}_{-0.000020}$\\
        $a_{\mathrm{pl}}$ [$\mathrm{AU}$]  & $0.7478^{+0.0056}_{-0.0060}$ & $0.7408^{+0.0073}_{-0.0082}$ & $0.7488^{+0.0046}_{-0.0057}$ & $0.7388^{+0.0070}_{-0.0080} $\\
        \hline
        Other parameters&&\\
        $V_{\mathrm{sys,A}}$ [$\mathrm{km~s^{-1}}$]  & $25.9867\pm0.0047$ & $26.0254^{+0.0056}_{-0.0058}$ & $26.0146\pm0.0045$ & $26.0531\pm0.0057$\\ 
        $V_{\mathrm{sys,B}}$ [$\mathrm{km~s^{-1}}$] & $26.0459\pm0.0060$  & $26.0970^{+0.0088}_{-0.0090}$ & $26.0639^{+0.0058}_{-0.0060}$  & $26.1144^{+0.0087}_{-0.0089}$\\ 
        \hline
    \end{tabular}
\end{table*}

\begin{figure*}
\includegraphics[width=0.475\textwidth]{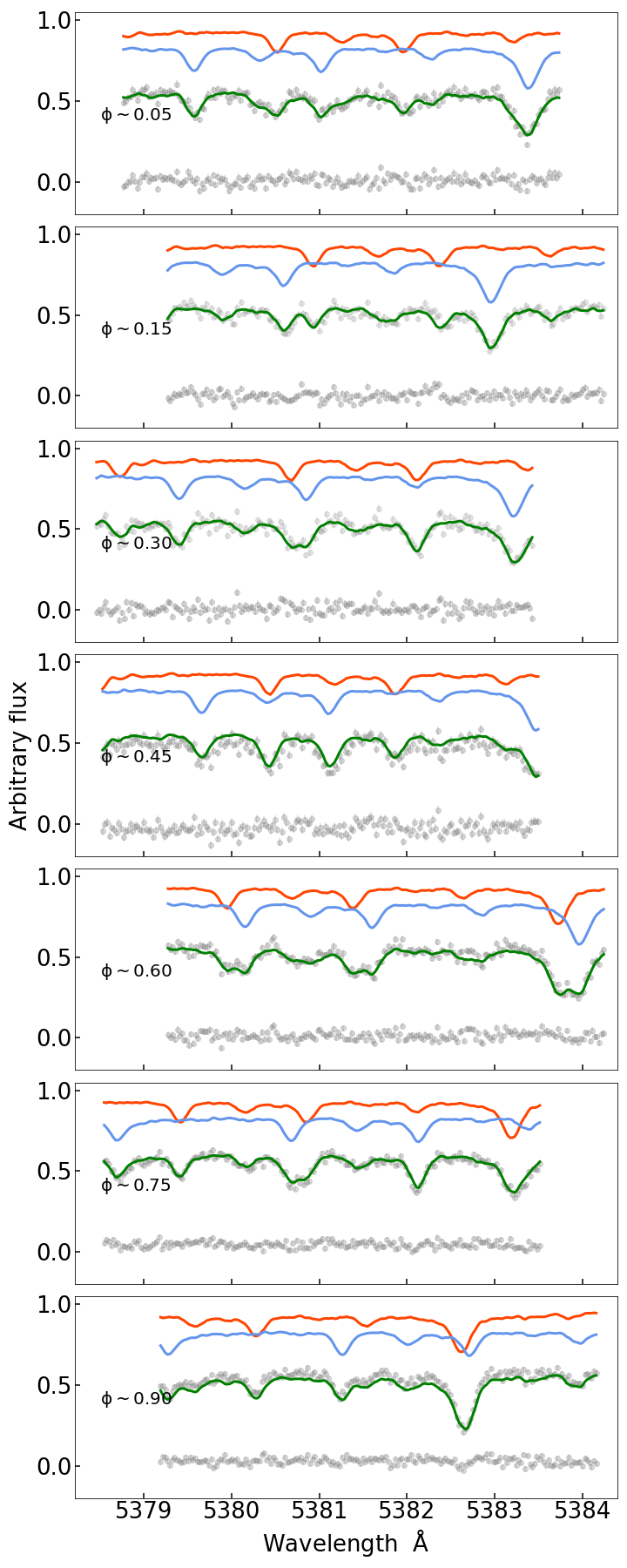}
\includegraphics[width=0.475\textwidth]{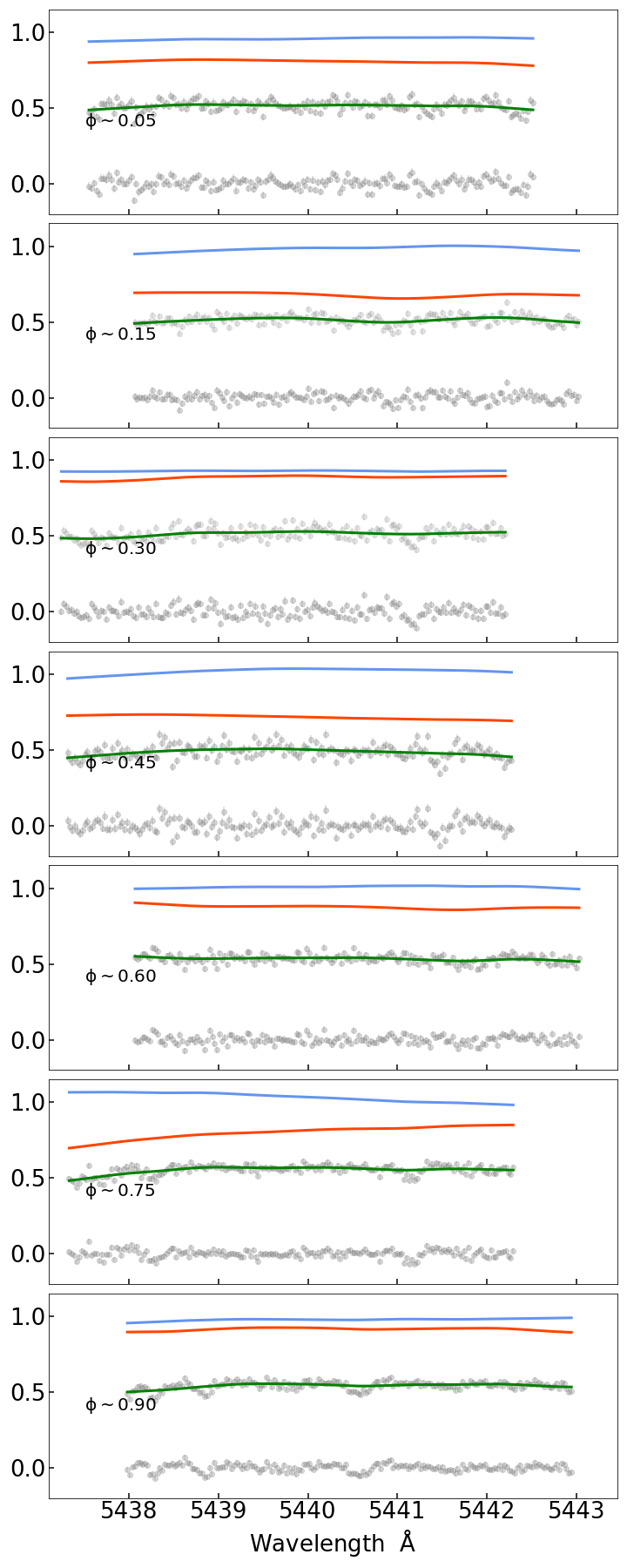}
\caption{Seven epochs of composite spectra for two chunks between 5379-5384$\Angstrom$ and 5437-5443 $\Angstrom$ are shown in grey with composite model in green. The mean realisation drawn from posterior predictive distribution are show in red and blue for the primary and secondary, respectively. The residuals are shown at the bottom of each panel. The first chunk shows several lines and the second chunk shows no lines in the composite spectra. The redial velocities from such chunks are weighted.}
\label{fig:sdphase}
\end{figure*}

\begin{figure*}
\includegraphics[width=1.03\textwidth]{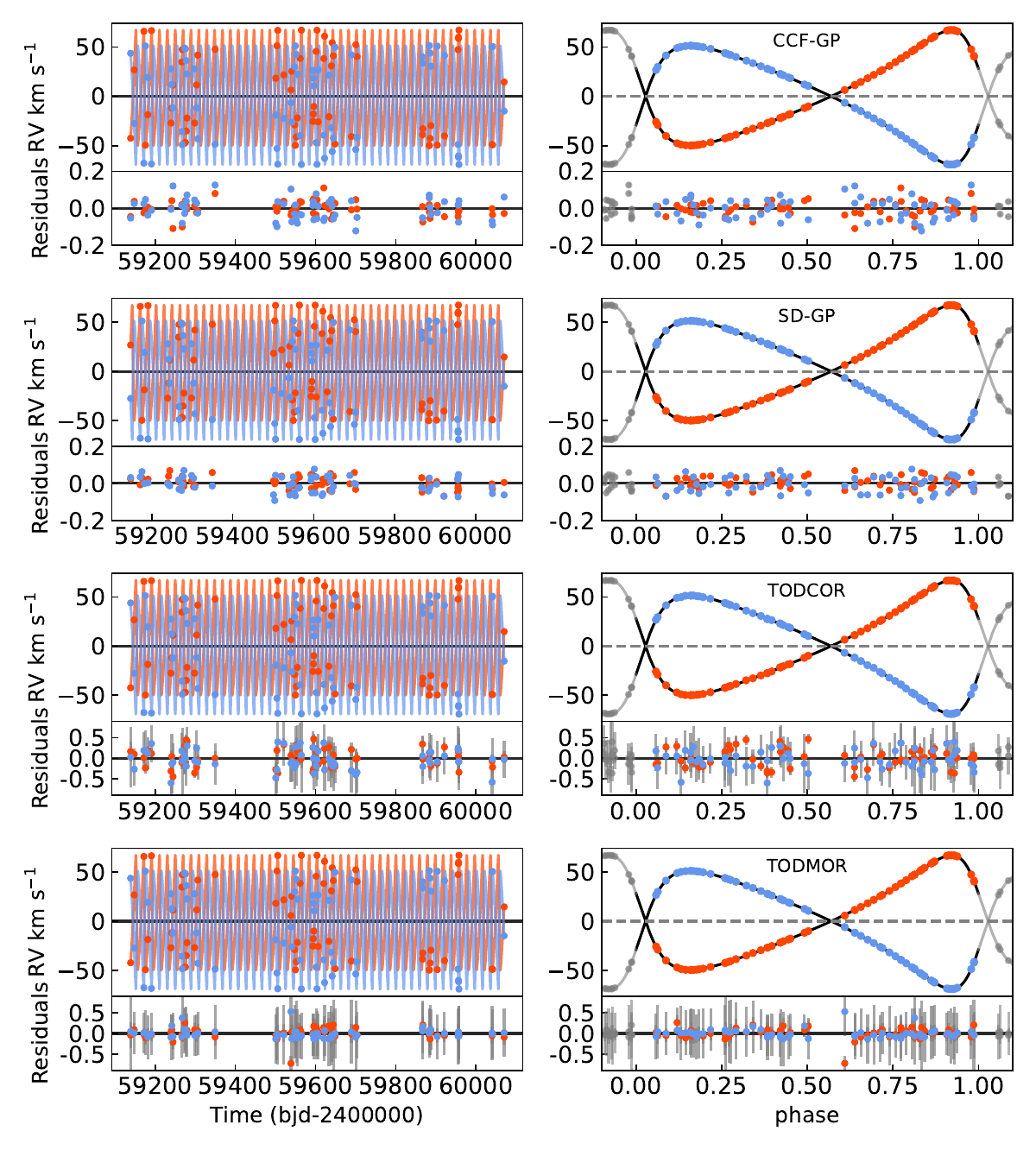}
\caption{SOPHIE radial velocities as a function of time and the corresponding residuals for the method 1 (SD-GP), method 2 (CCF-GP), TODCOR and TODMOR (top to bottom). The overplotted magenta and orange curves are the medial value of the posterior distribution models for the primary and the secondary stars corresponding to 2,000 random MCMC steps.  }
\label{fig:compmod}
\end{figure*}

\begin{figure}
\includegraphics[width=0.49\textwidth]{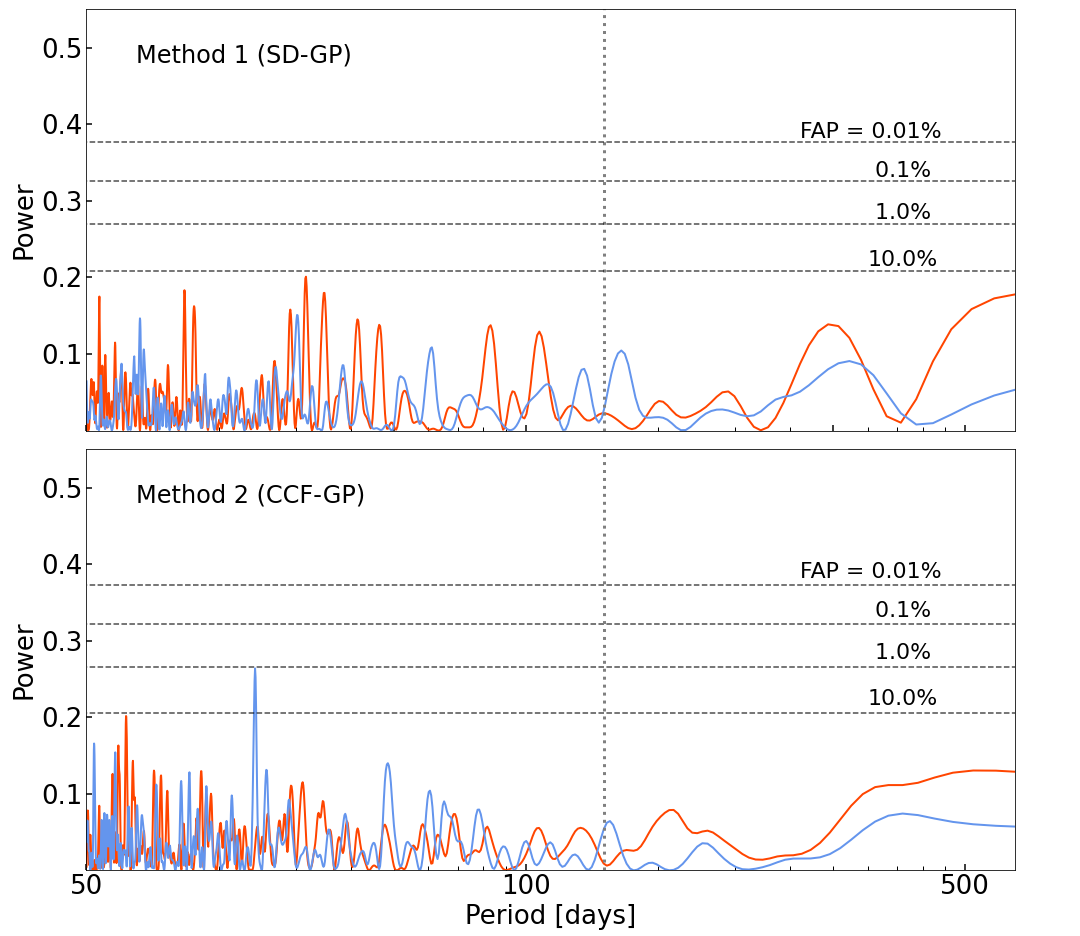}
\caption{The residual periodgram after removing the signal P$_b$ for both method 1 (top panel) and method 2 (bottom panel). The red and blue curves represent the primary and the secondary components of the binary. The grey vertical line indicates the period of $P_{\rm b}$.}
\label{fig:res_period}
\end{figure}

\begin{figure*}
\includegraphics[width=0.475\textwidth]{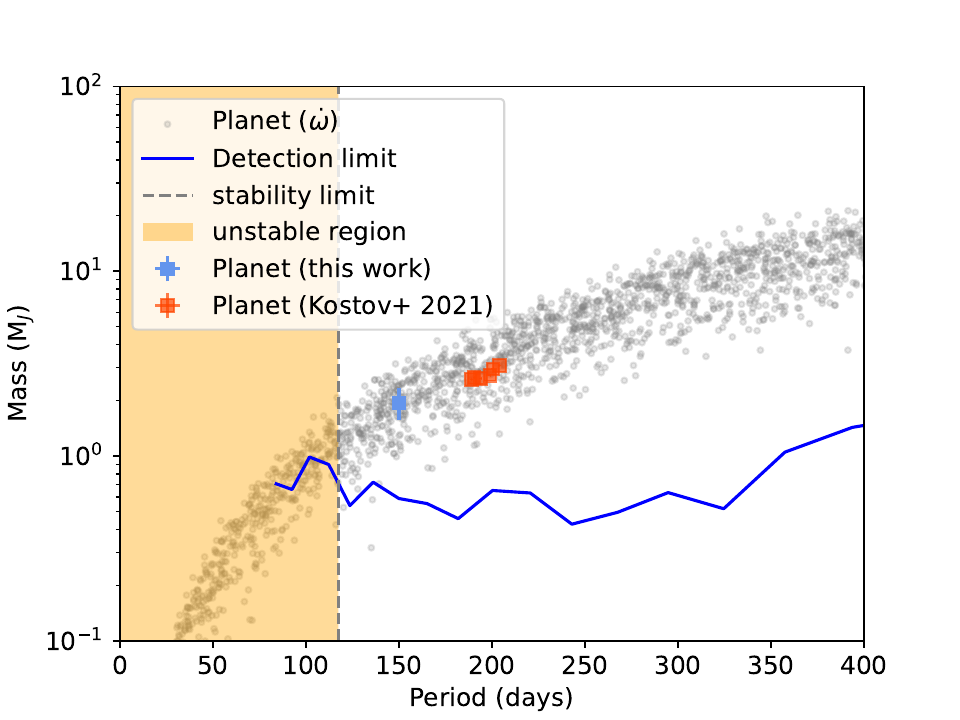}
\includegraphics[width=0.475\textwidth]{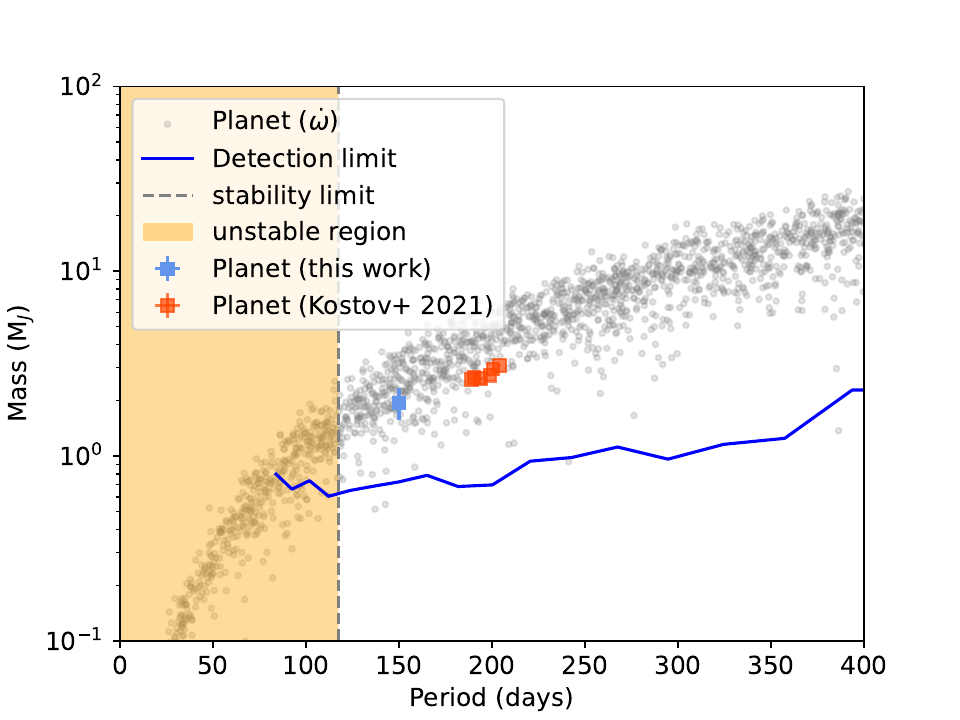}
\caption{Left panel method 1 (SD-GP) and right  panel method 2 (CCF-GP): Grey the scatter plot showing the planets consistent with the posterior distribution on the apsidal precession rate, the GR precession rate is accounted for both this and the third-body dynamical precession rate are calculated using the equations in \citet{Baycroft_2023}. Blue the locations of the solution presented in this work. Red the 6 suggested solutions from \citet{Kostov2021}. Dashed line the HW-stability limit. Blue line the detection limit with planet b removed.}
\label{fig:prec}
\end{figure*}

\begin{figure}
\includegraphics[width=0.49\textwidth]{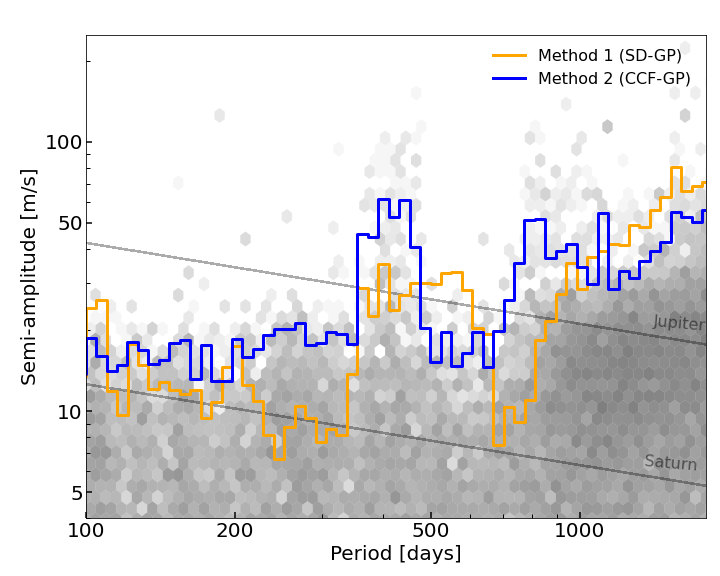}
\caption{The density of posterior samples from \texttt{kima} run on TIC172900988 with $N_{\rm p}$ fixed to 1. The orange and blue lines indicate the detection limits for method 1 and method 2.}
\label{fig:detlim_noplanet}
\end{figure}

\begin{figure}
\includegraphics[width=0.475\textwidth]{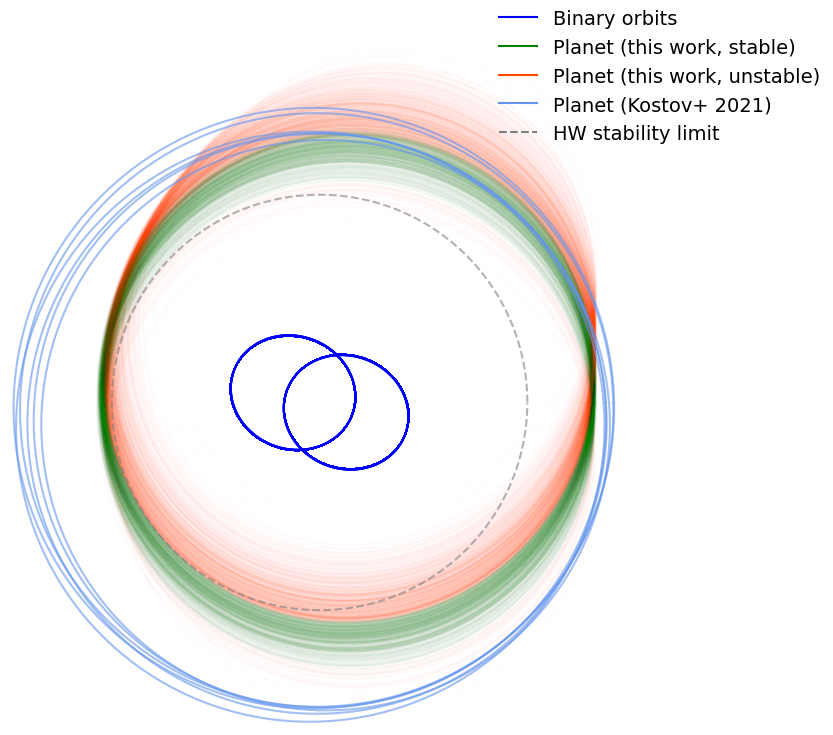}
\caption{Orbital configuration of TIC~172900988 showing the orbits of the binary and the planet. The red orbits are a random sample of 1000 posteriors from {\tt kima} fitting the radial velocities from SD-GP. The green orbits are the 6 suggested solutions from \citet{Kostov2021}, the dashed grey line is the stability limit as calculated by \citet{holman_1999}. The radial velocity data are fit along with the 1-2 punch transit times.}
\label{fig:orbit_view_12_punch}
\end{figure}

\label{lastpage}
\end{document}